\documentclass[12pt]{article}

\usepackage[margin=1in]{geometry}
\usepackage{amsfonts, amsmath, amssymb}
\usepackage[none]{hyphenat}
\usepackage{fancyhdr}
\usepackage{graphicx}
\usepackage{float}
\usepackage{graphicx}
\usepackage{framed} 

\usepackage{array}
\usepackage{lscape} 

\usepackage{listings} 
\usepackage{longtable}

\begin{document}

\begin{titlepage}
\begin{center}

\vspace*{1cm}
\Large{\textbf{}}\\
\vfill
\line(1,0){450}\\
\Large{\textbf{Prediction intervals for overdispersed Poisson data and their 
application in medical and pre-clinical quality control}}\\

\line(1,0){450}\\
\vfill

\flushleft
\small
Max Menssen$^{1\ast}$, Martina Dammann$^2$, Firas Fneish$^3$, David Ellenberger$^3$,
Frank Schaarschmidt$^1$ \\

$\ast$: Corresponding author\\

1: Department of Biostatistics, Leibniz University Hannover \\
2: BASF SE, Ludwigshafen Am Rhein, Germany \\
3: MS Forschungs- und Projektentwicklungs-gGmbH (MS Research and Projectdevelopment gGmbH [MSFP])\\

\end{center}
\end{titlepage}

\setcounter{page}{1}


\section*{Abstract}

In pre-clinical and medical quality control, it is of interest to assess the stability 
of the process under monitoring or to validate a current observation using historical
control data. Classically, this is done by the application of historical
control limits (HCL) graphically displayed in control charts. In many applications, HCL
are applied to count data, e.g. the number of revertant colonies (Ames assay)
or the number of relapses per multiple sclerosis patient. 
Count data may be overdispersed, can be heavily right-skewed and clusters may 
differ in cluster size or other baseline quantities (e.g. number of petri dishes 
per control group or different length of monitoring times per patient). \\
Based on the quasi-Poisson assumption or the negative-binomial distribution, we 
propose prediction intervals for overdispersed count data to be used as HCL. Variable
baseline quantities are accounted for by offsets. Furthermore, we provide a bootstrap 
calibration algorithm that accounts for the skewed distribution and achieves equal
tail probabilities. \\
Comprehensive Monte-Carlo simulations assessing the coverage probabilities of 
eight different methods for HCL calculation reveal, that the bootstrap calibrated
prediction intervals control the type-1-error best. Heuristics traditionally 
used in control charts (e.g. the limits in Sheward c- or u-charts or the mean $\pm$ 2 SD)
fail to control a pre-specified coverage probability. \\
The application of HCL is demonstrated based on data from the Ames assay and for
numbers of relapses of multiple sclerosis patients. The proposed prediction 
intervals and the algorithm for bootstrap calibration are publicly available via
the R package predint. \\

\textit{Keywords}: Bootstrap-calibration, Ames-test, negative-binomial distribution,
quasi-likelihood, Sheward control chart, historical control data

\newpage

\section{Introduction}

In pre-clinical research it is common sense, that a treatment of interest
(e.g. a new drug candidate) is labeled to be effective, if the response obtained in treated individuals 
(e.g. patients or model organisms) differs significantly from the response obtained in a concurrent
(negative) control. 
In pre-clinical risk assessment, such as toxicological studies, the experimental design
usually contains a negative control group, several groups that received increasing
dosages of the compound of interest and sometimes at least one 
positive control in order to show the proficiency of the assay in use (Hothorn 2015, OECD 489). \\
It frequently happens, that several studies are run which explore the impact of
\textit{different} treatments 
(e.g. different drug candidates) on the \textit{same} endpoint. If this is the case, one can 
exploit the observed data from historical control groups - the so called historical 
control data (HCD) - in order to calculate control limits for the validation of 
a recent (or future) control group of a study on the same endpoint (Menssen 2023, 
Dertinger et al. 2023, Kluxen et al. 2021). \\
In medical quality control, the monitoring of adverse events such as
the number of pressure ulcers per patient obtained over a certain time period in 
a certain ICU (Still et al. 2013) is of highest interest.
In this context the number of adverse events of different patients is tracked over 
time and used in order to validate, if the number of adverse events in a set of other patients
is in line with the historical data e.g. by the application of control limits 
(Koetsier et al. 2012, Chen et al. 2010). Hence, this type of data is collected for similar reasons 
as the HCD in toxicology.\\
It has to be stressed that the application of HCD depends on the strong, but necessary,
assumption that the HCD as well as current (or future) observations are derived 
from the \textit{same} data generating process and therefore, are exchangeable 
(Menssen 2023, Menssen and Schaarschmidt 2022, Menssen and Schaarschmidt 2019,
Gsteiger et al. 2013). Hence, this matter and its impact on the use and compilation 
of HCD is widely discussed (Menssen 2023, Dertinger et al. 2023, Coja et al. 2022, 
Kluxen et al. 2021, Viele et al. 2014) and several regulatory guidelines and other publications
directly refer to this topic (Gurjanov et al. 2023, EU commission regulation 283/2013,
Hayashi et al. 2011, Greim et al. 2003). Due to the fact that several guidelines
(e.g. OECD 471, OECD 490) explicitly call for the presentation of HCD along with 
the outcome of the current study, most laboratories maintain their own historical 
control data base and efforts are made to share and report HCD across organizations 
e.g. via the NTP historical control data base (NTP 2024), the RITA data base
(Deschl et al. 2002) or eTransafe project (Pognan et al. 2021). \\
In medical quality control, the graphical display of historical control limits (HCL) 
in control charts is widely discussed and different types of control charts are 
in use (Sachlas et al. 2019, Koetsier et al. 2012, Lyren et al. 2017, 
Benoit et al. 2019). Despite the fact, that
the OECD recommends the application of control charts for several years (OECD 2017),
their application does not play a major role in toxicology so far. Anyhow, this 
topic was recently discussed on the International Workshop on Genotoxicity Testing 
(IWGT) and Dertinger et al. 2023 provided examples on the use of control charts 
in order to assess the quality of HCD obtained from different genotoxicity 
assays. \\
The different applications of HCL have in common, that the desired control limits are 
calculated in order to evaluate, if certain observation(s) (historical, current or future) 
belong to the central $100(1- \alpha)$\% of the underlying distribution (usually 95\% or 99.7\%) 
or if they can be treated as "outliers". Over the past decades several heuristic
methods were applied for the calculation of HCL e.g. the mean $\pm$ 2 standard deviations
or the control limits applied in classical Sheward control charts. Furthermore,
several authors proposed the application of prediction intervals in this context,
since they should directly converge against the lower $\alpha /2$ 
and the upper $1- \alpha/2$ quantiles of the underlying distribution. The calculation
of HCL based on prediction intervals was proposed in order to validate the concurrent
(negative) control in toxicity or carcinogenicity assays (Menssen 2023, Menssen 
and Schaarschmidt 2019, Kluxen et al. 2021, Dertinger et al. 2023), in the context 
of anti-drug anti-body detection (Francq et al. 2019, Menssen and Schaarschmidt 2022, 
Hoffman and Berger 2001, Schaarschmidt et al. 2015) or in the context of medical
control charts (Chen et al. 2011).  \\
Most of the work regarding prediction intervals that are aimed to serve as HCL was done for 
observations that are continuous and hence are assumed to follow at least approximately
a (multivariate) normal distribution (e.g. in the context of anti-drug anti-body cut points) 
(Franqc et al. 2019, Menssen and Schaarschmidt 2022) or 
for binary observations such as the number of rats with a tumor vs. the number
of rats without a tumor (Menssen and Schaarschmidt 2019) or for the cumulative 
sum of binomial proportions (Chen et al. 2011). Contrary, the application of 
prediction intervals for count data that match the clustered data structure of 
toxicological or medical HCD has received less attention so far.\\
Classically, count data is modeled based on the Poisson distribution and several
non-heuristic prediction intervals for this assumption are reviewed in Meeker et al. 2017.
However, it has to be stressed, that both, the control limits in Sheward c- and 
u-charts, classically applied to count data in quality control,
as well as the prediction intervals given by Meeker et al. 2017 are based 
on the assumption, that the historical and the current observations are independent
realizations of the \textit{same} Poisson process.\\
The assumption of independent and identically distributed observations might be 
sensible in industrial quality control, in which the number of nonconforming products 
per production unit is monitored over time. But, medical or toxicological HCD 
usually follows a hierarchical design in which certain individuals are nested 
in a certain control group or health care unit. 
Since several factors such as the genetic condition of patients or personnel
between control groups or patients can change, it is likely that the observations within a 
certain control or individual (e.g. patient) are positively correlated (Menssen 2023, Menssen and Schaarschmidt 2019, 
McCullagh and Nelder 1989, Demetrio et al. 2014). This results in observations that show higher
variability than possible under the simple Poisson distribution. Usually, this 
effect is called overdispersion or extra-Poisson variability and its presence
can be expected in biological data (McCullagh and Nelder 1989, Demetrio et al. 2014).
Hence, this manuscript is aimed to provide methodology for the calculation of 
prediction intervals for overdispersed count data which can be applied in 
two ways: The validation of a current or future observation based on HCD as
well as for the assessment of the quality and stability of HCD using improved versions 
of Sheward c- and u-charts. \\
The manuscript is organized as follows: 
The next section outlines two common models for overdispersed data. 
Among heuristic methodology, section \ref{sec::control_limits} introduces methods
for the calculation of prediction intervals for overdispersed observations.
Section \ref{sec::real_life_data} gives an overview about real life data with 
a toxicological or medical background.
Section \ref{sec:simulation} provides simulated coverage probabilities for each of the
methods provided in section \ref{sec::control_limits}.
The application of the proposed methods is demonstrated in section \ref{sec::application}.
The last two sections provide a discussion and conclusions.


\section{Models for overdispersed Poisson data} \label{sec:Model}

Modeling of count data is usually done based on the Poisson distribution, assuming that
\begin{gather}
	Y \sim Pois(\lambda) \label{eq:simple_poisson} \\
	E(Y) = var(Y) = \lambda \notag
\end{gather}
with $\lambda$ as the Poisson mean and variance and $Y$ as the Poisson distributed 
random variable. But, this approach ignores the clustered structure of medical and
toxicological HCD. In toxicology, HCD is usually comprised of $h=1, 2, \ldots H$ 
historical control groups of which each contains $i= 1, 2, \ldots, n_h$ experimental 
units (e.g. $n_h$ petri dishes per control group). Similarly, medical HCD is usually 
comprised of $h=1, 2, \ldots H$ patients for which the number of adverse events
is counted during the time interval $n_h$ each patient spend under monitoring 
(e.g. $n_h$ days or years). Generally spoken, $h=1, 2, \ldots H$ is the index for 
the historical clusters, regardless what these clusters are comprized of
(single patients, control groups or whatever).\\
In this case, it is a common strategy to model the total number of observations 
per cluster $Y_h$ which represents the sum of all observations per control group 
or patient over their corresponding $n_h$ experimental units or time intervals.
Since $n_h$ is not necessarily a constant, it has to enter the model as 
an offset, such that the expectation for the numbers of observations remains 
constant in the case were $n_h=1$ 
\begin{gather}
	E(Y_{ih})=\lambda = \frac{\lambda_h}{n_h}
\end{gather}
with $\lambda_h$ as the expectation for the total number of observations $Y_h$
observed over $n_h$ experimental units or time intervals
\begin{gather}
	E(Y_h) = var(Y_h) = \lambda_h = n_h \lambda. \label{eq:simple_poisson_offset}
\end{gather}
Further details on the modeling of Poisson type data using offsets 
are given in the supplementary material.\\
This clustered structure gives rise to possible overdispersion, meaning that the 
variability of the observed data exceeds the variability of a simple Poisson
random variable. This can be caused by positive correlations between the observations
in each cluster (Demetrio et al. 2014, McCullagh and Nelder 1989). In the context of
the Ames test (OECD 471) this would mean, that the observations 
\textit{within} each control group might descent from the same bacteria stock, operated by possibly 
the same personnel. But, bacteria stocks, personnel and maybe other conditions might randomly change 
\textit{between} the control groups of different historical studies. It is 
obvious that the same principle applies also for medical HCD that is 
comprised of different patients with different genetic conditions which are possibly
cared by different personnel. \\
A common way to model overdispersion is the application of a generalized linear model
(GLM) that either depend on the quasi-likelihood approach (quasi-Poisson assumption)
or that is based on the assumption that the observations follow a negative-binomial
distribution.\\
In the quasi-Poisson approach it is assumed that the variance is 
inflated by a dispersion parameter $\phi$ that is constant for all clusters,
such that
\begin{gather}
	var(Y_{h}) =  \phi n_h \lambda \label{eq:varQB} 
\end{gather}
with $E(Y_h) = n_h \lambda$ and $\phi > 1$.\\
If the data is assumed to follow a negative-binomial distribution, the cluster means 
$\lambda_h$ itself are assumed to follow a gamma distribution with parameters 
$a=1/\kappa$ and $b_h=1/(\kappa n_h \lambda)$, expected value $E(\lambda_h)=a/b_h = n_h \lambda$
and variance $var(\lambda_h)=a/b_h^2=\kappa n_h^2 \lambda^2$
\begin{gather*}
	\lambda_h \sim gamma(a, b_h) \\
	Y_h \sim Pois(\lambda_h)
\end{gather*}
with
\begin{gather}
	E(Y_h)=n_h \lambda \notag \\
	var(Y_h)=n_h \lambda + \kappa n_h^2 \lambda^2 = n_h \lambda (1+ \kappa n_h \lambda) \label{eq:varBB}
\end{gather}
with $\kappa > 0$. Further details about this model are given in section 1 of the supplementary 
materials. Please note, that the quasi-Poisson assumption
and the negative-binomial distribution are not in contradiction to each other, 
if all offsets are the same, such that $n_1 = n_2 = \ldots = n_H = n$, because 
in this case the part of the negative-binomial variance that contributes to the 
overdispersion $(1+ \kappa n \lambda)$ is constantly inflating 
the Poisson variance, similarly to the quasi-Poisson dispersion parameter $\phi$. 

\section{Historical control limits for overdispersed Poisson data} \label{sec::control_limits}

All historical control limits $[l, u]$ mentioned below are calculated based on 
observed events $y_h$ counted over $n_h$ historical experimental units 
(or time intervals) and are aimed to cover a certain number of observations $y^*$ 
counted over $n^*$ units of the offset variable (e.g. no. of petri dishes or the 
monitoring time of patients) with coverage probability 
\begin{equation}
        P(l \leq y^* \leq u) = 1-\alpha.
\end{equation}
Hence, they are aimed to approximate the central $100(1-\alpha) \%$ of the underlying
distribution.
But, overdispersed count data becomes heavily right skewed with a decreasing Poisson
mean and / or with an increasing amount of overdispersion (see supplementary materials).
Therefore, it is crucial to ensure that the desired control limits account for equal tail 
probabilities in a way that
\begin{equation}
        P(l \leq y^*) = P(y^* \leq u) = 1-\alpha/2. \label{eq::equal_tails}
\end{equation}
If this is the case, the desired control limits converge against the true $\alpha/2$
and $1-\alpha/2$ quantiles of the underlying distribution of $y^*$ and hence properly
approximate its desired central $100(1-\alpha)\%$.


\subsection{Heuristical HCL for count data} \label{sec::heristics}

One heuristic method for the calculation of HCL which is frequently applied 
in toxicology is
\begin{gather}
        [l,u]=\bar{y} \pm k SD \label{eq:mean_pm_se}
\end{gather}
with $\bar{y}=\sum_{h=1}^H y_h/H$ as the sample mean, $SD=\sqrt{\frac{\sum_{h=1}^H(y_h - \bar{y})^2}{H-1}}$ as the sample standard deviation, $H$ as the total number of clusters (e.g. 
control groups or patients) and $k$ as the factor that determines the 
desired coverage probability (Menssen 2023, Kluxen et al. 2021, Levy et al. 2019, 
Rotolo et al. 2021, Prato et al. 2023). 
In toxicology, $k$ is usually set to 2, in order to set the nominal 
coverage probability to 95.4 \%. In quality control HCL are classically 
calculated based on $k=3$ in order to cover an observation with 99.7 \% coverage probability 
(Montgomery 2020). Note that the mean $\pm$ k SD interval 
is based on a simple normal approximation which lacks an explicit assumption 
about the mean-variance relationship and hence, heuristically accounts for 
overdispersion (and underdispersion as well). However, the application of the 
mean $\pm$ k SD is explicitly based on the assumption, that all observations have the
same variance and that the underlying distribution can be satisfactorily approximated
by a normal distribution. For count data, this is only the case, if all offsets are the same
$n_h = n_{h'} = n^*$. Consequently its application to right-skewed count data that depends on 
different offsets should be avoided.\\
The control limits typically used in a Sheward c-chart are given as 
\begin{gather}
        [l, u]=  \bar{y} \pm k \sqrt{\bar{y}}.
\end{gather}
with $\bar{y}=\frac{\sum_{h=1}^H y_h}{H}$.
Similar to the mean $\pm$ k SD control limits, the limits in a c-chart do not 
account for different offsets ($n_h$) and hence, are only applicable if all offsets
are the same ($n_h = n_{h'} = n^*$).\\
Methodology to set control limits in the case of different offsets, is given by the
Sheward u-chart
\begin{gather}
        [l, u]=  \bar{u} \pm k \sqrt{\frac{\bar{u}}{n^*}}
\end{gather}
with $\bar{u}=\frac{\sum_{h=1}^H u_h}{H}$, $u_h=y_h/n_h$ and $n^*$ as the offset
attached to the prediction (e.g. no. of petri dishes in the curent control group or
monitoring time of a certain patient) which can differ from the historical offsets
$n_h$. \\
Since the control limits in c- and u-charts are explicitly based on a normal approximation
of the Poisson 
distribution, it is assumed that the mean equals the variance (equations
\ref{eq:simple_poisson} and \ref{eq:simple_poisson_offset}). Therefore, these 
two types of control limits do not account for overdispersion. \\
In order to overcome this problem, Laney 2006 proposed a version of the u-chart 
that corrects for between study overdispersion in a way that the between study 
overdispersion is inflating the Poisson variance as a constant (quasi-Poisson 
assumption). Following Mohammed and Laney 2006, the control limits for an overdispersion
corrected u-chart are given as
\begin{gather}
        [l, u]=  \bar{u} \pm k \sqrt{\frac{\bar{u}}{n^*}} \sqrt{\frac{\sum_h{(z_h - \bar{z})^2}}{H}}
\end{gather}
with $z_h=\frac{u_h - \bar{u}}{\sqrt{\bar{u}/n_h}}$ and $\bar{z} = \frac{\sum_h z_h}{H}$.  \\
Anyhow, all four methods have two major drawbacks for practical application: 
They do not account for the uncertainty of the estimates for the model parameters
and hence, should yield control limits that are too narrow to approximate the 
desired percentage in the center of the underlying distribution (especially if the number of
historical observations is low). Furthermore, they are symmetrical around the mean, 
but overdispersed count data can be heavily right-skewed. Hence, they do not
ensure that both interval borders cover a future observation with the same probability
as required in equation \ref{eq::equal_tails}.

\subsection{Prediction intervals for overdispersed count data}
Several methods for the calculation of prediction intervals based on one unstructured 
sample of independent and identically Poisson distributed observations $y$ 
following the model given in equation \ref{eq:simple_poisson_offset} are reviewed
in standard text books regarding statistical intervals (Hahn and Meeker 1991,
Meeker et al. 2017). 
The prediction intervals for overdispersed Poisson data that are introduced below,
are based on asymptotic methodology proposed by Nelson 1982. Following this approach, 
an asymptotic prediction interval $[l, u]$ can be computed based on one historical 
sample of observations $y$ (e.g. obtained from one single control group or patient)
obtained over the offset $n$ (e.g. no. of experimental units in one historical
control group or the monitoring time of one historical patient). This prediction 
interval aims to cover one further random realization $y^*$ that is obtained over 
its corresponding offset $n^*$ (e.g. no. of experimental units in one current
control group or the monitoring time of one further patient) with nominal coverage
probability 
\begin{equation*}
	P(l < y^* < u) = 1-\alpha.
\end{equation*}
This interval is based on the assumption that 
\begin{equation*}
	\frac{\hat{y}^* - Y^*}{\sqrt{\widehat{var}(\hat{y}^* - Y^*)}} = \frac{n^* \hat{\lambda} - Y^*}{\sqrt{\widehat{var} \big({n^* \hat{\lambda} - Y^* \big)}}} = \frac{n^* \hat{\lambda} - Y^*}{\sqrt{\widehat{var}({n^* \hat{\lambda}) + \widehat{var}(Y^*)}}}  
\end{equation*}
is approximately standard normal. \\
In this notation $\hat{\lambda}$ is 
the estimate for the Poisson mean obtained from the historical observations.
The standard error of the prediction is 
\begin{equation*}
	\widehat{se}(\hat{y}^* - Y^*)=
	\sqrt{\widehat{var}({n^* \hat{\lambda}) + \widehat{var}(Y^*)}} = 
	\sqrt{\frac{n^{*2} \hat{\lambda}}{n} + n^* \hat{\lambda}}
\end{equation*} 
and $n$ is the offset the historical number of counted observations $y$ is based on.
The corresponding asymptotic prediction interval is given by
\begin{equation}
        [l,u]=n^* \hat{\lambda} \pm z_{1-\alpha/2} \sqrt{\frac{n^{*2} \hat{\lambda}}{n} + n^* \hat{\lambda}} \label{eq:simple_pi}
\end{equation} 
However, the prediction interval given in eq. \ref{eq:simple_pi} is based on the
assumption of independent observations obtained in one single cluster (e.g. patient or control group)
and hence, does not care for the clustered structure usually found in historical control
data. Consequently, this prediction interval does not account for possible overdispersion
and hence, was adapted to the clustering in a way that possible overdispersion is 
taken into account. This was done by the adaption of the formulas for the variance
estimates that define the prediction standard error $\widehat{se}(\hat{y}^* - Y^*) =
\sqrt{\widehat{var}({n^* \hat{\lambda}) + \widehat{var}(Y^*)}}$.\\
If overdispersion is modeled based on the quasi-Poisson assumption, the prediction 
standard error becomes $\widehat{se}(\hat{y}^* - Y^*) =\sqrt{\frac{n^{*2} \hat{\phi} \hat{\lambda}}{\bar{n} H} + n^* \hat{\phi} \hat{\lambda}}$
and the corresponding prediction interval is given by
\begin{equation}
	[l,u]=n^* \hat{\lambda} \pm z_{1-\alpha/2} \sqrt{\frac{n^{*2} \hat{\phi} \hat{\lambda}}{\bar{n} H} +
	n^* \hat{\phi} \hat{\lambda}} \label{eq::quasi_pois_pi}
\end{equation}
with $\hat{\phi} \geq 1$, $\bar{n}=\frac{\sum_h^H n_h}{H}$ and $n_h$ as the offsets
(e.g. number of petri dishes per historical control group) in $h=1, 2, \ldots, H$ 
historical clusters (e.g. control groups or patients). \\
Similarly, a prediction interval that is based on the negative-binomial distribution is given by
\begin{equation}
	[l,u]=n^* \hat{\lambda} \pm z_{1-\alpha/2} 
	\sqrt{n^{*2} \frac{\hat{\lambda} + \hat{\kappa} \bar{n} \hat{\lambda}}{\bar{n} H} + 
	(n^* \hat{\lambda} + \hat{\kappa} n^{*2} \hat{\lambda}^2)}. \label{eq::neg_bin_pi}
\end{equation}
with $\widehat{se}(\hat{y}^* - Y^*) = \sqrt{n^{*2} \frac{\hat{\lambda} + 
\hat{\kappa} \bar{n} \hat{\lambda}}{\bar{n} H} + (n^* \hat{\lambda} + 
\hat{\kappa} n^{*2} \hat{\lambda}^2)}$.
Further details on the derivation of the prediction variances for both intervals 
that account for overdispersion are given in section 3 of the supplementary material.\\

\subsection{Bootstrap calibration}
As mentioned above, overdispersed Poisson data can be heavily right-skewed. Therefore,
it is crucial to ensure equal tail probabilities of the applied prediction intervals 
(see equation \ref{eq::equal_tails}). Consequently, the 
applied methodology has to enable the calculation of asymmetrical prediction 
intervals. This is ensured by the application of a bootstrap calibration procedure 
in which both interval limits are calibrated individually using the algorithm given
in the box below.

\begin{framed}
\textbf{Bootstrap calibration of the proposed prediction intervals}
\begin{enumerate}
        \item Based on the historical data $\boldsymbol{y}$ find
        estimates for the model parameters $\hat{\boldsymbol{\theta}}$, with 
        $\hat{\boldsymbol{\theta}}=(\hat{\lambda}, \hat{\phi})$ in the quasi-Poisson 
        case and $\hat{\boldsymbol{\theta}}=(\hat{\lambda}, \hat{\kappa})$ in the
        negative-binomial case
        \item Based on $\hat{\boldsymbol{\theta}}$, sample $B$ parametric bootstrap 
        samples $\boldsymbol{y}_b$ following the same experimental design as the 
        historical data (for sampling algorithms see section 4 of the supplementary material)
        \item Draw $B$ further bootstrap samples $y^*_b$ following the 
        same experimental design as the current observations (e.g. the same offset $n^*$)
        \item Fit the initial model to $\boldsymbol{y}_b$ in order to obtain 
        $\hat{\boldsymbol{\theta}}_b$
        \item Based on $\hat{\boldsymbol{\theta}}_b$, calculate $\widehat{var}_b(n^* \hat{\lambda})$
        and $\widehat{var}_b(Y^*)$
        \item Calculate lower prediction borders $l_b = n^* \hat{\lambda}_b - q_l 
        \sqrt{\widehat{var}_b(n^* \hat{\lambda}) + \widehat{var}_b(Y^*)}$. Note that
        all $l_b$ depend on the same value for $q_l $.
        \item Calculate the bootstrapped coverage probability 
        $\hat{\psi}_l=\sum_b I_b \text{ with } 
        I_b = 1 \text{ if } l_b \leq y^*_b \text{ and }
        I_b = 0 \text{ if } y^*_b < l_b$
        \item Alternate $q_l$ until $\hat{\psi}_l \in (1-\frac{\alpha}{2}) \pm t$ 
        with $t$ as a predefined tolerance around $1-\frac{\alpha}{2}$
        \item Repeat steps 5-7 for the upper prediction border with 
        $\hat{\psi}_u=\sum_b I_b \text{ with } 
        I_b = 1 \text{ if } y^*_b \leq u_b \text{ and }
        I_b = 0 \text{ if } u_b < y^*_b$
        \item Use the corresponding coefficients $q^{calib}_l$ and $q^{calib}_u$ 
        for interval calculation
        \begin{gather*}
                \big[l= n^* \hat{\lambda} - q^{calib}_l 
                \sqrt{\widehat{var}(n^* \hat{\lambda}) + \widehat{var}(Y^*)},\\
                \quad\quad\quad\quad\quad 
                u= n^* \hat{\lambda} + q^{calib}_u
                \sqrt{\widehat{var}(n^* \hat{\lambda}) + \widehat{var}(Y^*)} \big]
        \end{gather*}
\end{enumerate}
\end{framed}
This algorithm is a modified version of the bootstrap calibration approach of 
Menssen and Schaarschmidt 2022. The search for $q_L^{calib}$ and $q_U^{calib}$ 
in steps 7 and 8 is based on the same bisection procedure that was described by 
Menssen and Schaarschmidt 2022 using a tolerance of $t=0.001$.\\
The bootstrap calibrated version of the prediction interval that is based on the 
quasi-Poisson assumption is given by
\begin{gather}
	\Big[l=n^* \hat{\lambda} - q_l^{calib} \sqrt{\frac{n^{*2} \hat{\phi} \hat{\lambda}}{\bar{n} H} +
	n^* \hat{\phi} \hat{\lambda}}, \label{eq::QP_calib} \\
	\quad\quad\quad\quad\quad u=n^* \hat{\lambda} + q_u^{calib} \sqrt{\frac{n^{*2} \hat{\phi} \hat{\lambda}}{\bar{n} H} +
	n^* \hat{\phi} \hat{\lambda}} \Big]. \notag
\end{gather}
Similarly, the bootstrap calibrated negative-binomial prediction interval is given by
\begin{gather}
	\Big[l=n^* \hat{\lambda} - q_l^{calib} \sqrt{n^{*2} \frac{\hat{\lambda} + \hat{\kappa} \bar{n} \hat{\lambda}}{\bar{n} H} + 
	(n^* \hat{\lambda} + \hat{\kappa} n^{*2} \hat{\lambda}^2)} \label{eq::NB_calib}\\
	\quad\quad\quad\quad\quad u=n^* \hat{\lambda} + q_u^{calib} \sqrt{n^{*2} \frac{\hat{\lambda} + \hat{\kappa} \bar{n} \hat{\lambda}}{\bar{n} H} + 
	(n^* \hat{\lambda} + \hat{\kappa} n^{*2} \hat{\lambda}^2)} \Big]. \notag
\end{gather}

\subsection{Computational details and estimation}

The proposed prediction intervals are implemented in the R package predint (Menssen 2023).
Uncalibrated prediction intervals can be computed using the functions \texttt{qp\_pi()}
and \texttt{nb\_pi()}. The bootstrap
calibrated prediction intervals are implemented in the functions \texttt{quasi\_pois\_pi()}
and \texttt{neg\_bin\_pi()}. For the quasi-Poisson assumption, the estimates $\hat{\lambda}$
and $\hat{\phi}$ were obtained based on \texttt{stats::glm}. The estimates for
the negative-binomial distribution ($\hat{\lambda}$ and $\hat{\kappa}$) were 
obtained based on \texttt{MASS::glm.nb}.\\
The bisection procedure used for bootstrap calibration is implemented in the function
\texttt{bisection()}. This function takes three different lists that contain the bootstrapped 
expected observations $\hat{y}^*_b$, the bootstrapped standard errors $\sqrt{\widehat{var}(n^* \hat{\lambda})_b + \widehat{var}(Y^*)_b}$ and the bootstrapped further observations $y^*_b$ as input and
returns values for $q_l$ and $q_u$. Hence, this function enables the calculation of
bootstrap calibrated prediction intervals for a broad range of applications and 
data scenarios (given that a parametric model can be fit to the data from which
bootstrap samples can be derived).


\section{Properties of real life HCD} \label{sec::real_life_data}

\subsection{Pre-Clinical HCD from the Ames Test} \label{sec::ames_HCD}

\begin{figure}[h!]
  \includegraphics[width=\linewidth]{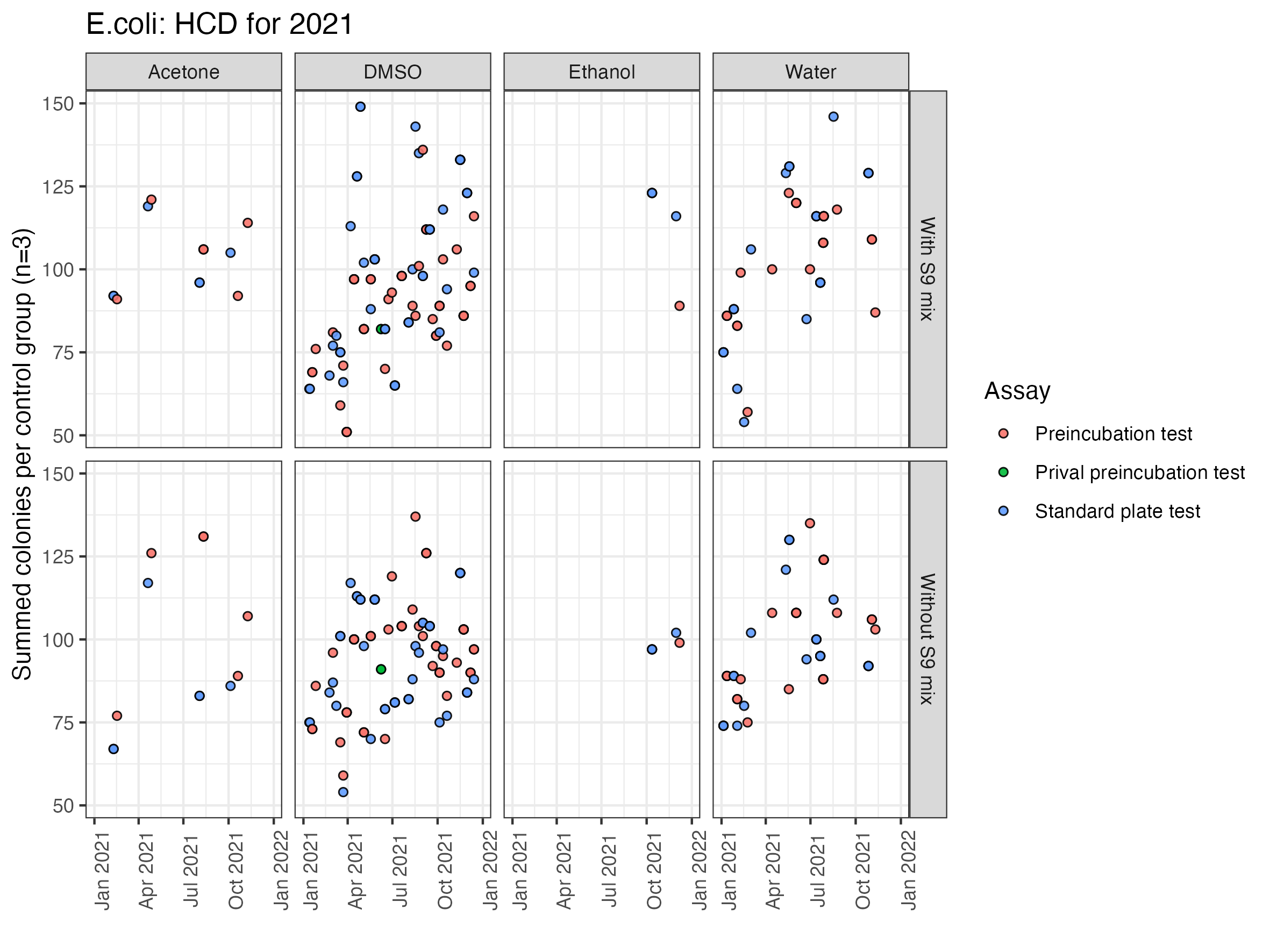}
  \caption{Overview about the number of revertant colonies in historical negative
  controls using E. coli (summed over three petri dishes per control group).}
  \label{fig:e_coli_hcd}
\end{figure}

This section gives an overview about a real life data base containing historical
negative control groups from the Ames test (OECD 471). Because its ability to 
score the mutagenic potential of a chemical compound and its conduction
is relatively low cost, the Ames test is one of the earliest tests in the test battery in
drug and pesticide development and is conducted by numerous laboratories and contract research 
organizations around the globe (Tejs 2008). \\
In the Ames test, bacteria 
strains that lack the ability to synthesize an essential amino acid are cultivated
on medium that lacks this particular amino acid. Hence, only colonies that stem from one 
single mutated bacterium are able to survive on
the medium (since they carry a mutation which enables the synthesis of the needed amino 
acid).\\
Usually, the experimental design is comprised of several treatment groups and a negative
control, each consisting of three petri dishes. If the number of 
revertants in the treatment groups has significantly increased compared to the untreated
negative control, the compound of interest is considered to be mutagenic and 
hence its development will not be pursued further.\\
The presented data base contains observations (counted number of revertant bacteria
colonies per petri dish) from 902 different control groups, each comprised of three petri dishes 
(except one that was comprised of two petri dishes). \\
These negative controls came from experiments run in 2021 
that differed with regard to the bacteria stems (E. coli, TA 98, TA 100, TA 1535
and TA 1537), the vehicles (acetone, DMSO, ethanol, water), experiment run with 
or without S9 mix and the type of the assay (preincubation test, 
prival preincubation test, standard plate test).
Hence, the data base contains 90 different data sets, of which each represents
another experimental setup. A graphical overview about the historical controls
for E. coli is given in figure \ref{fig:e_coli_hcd}.
 \\
 
\begin{figure}[h!]
  \includegraphics[width=\linewidth]{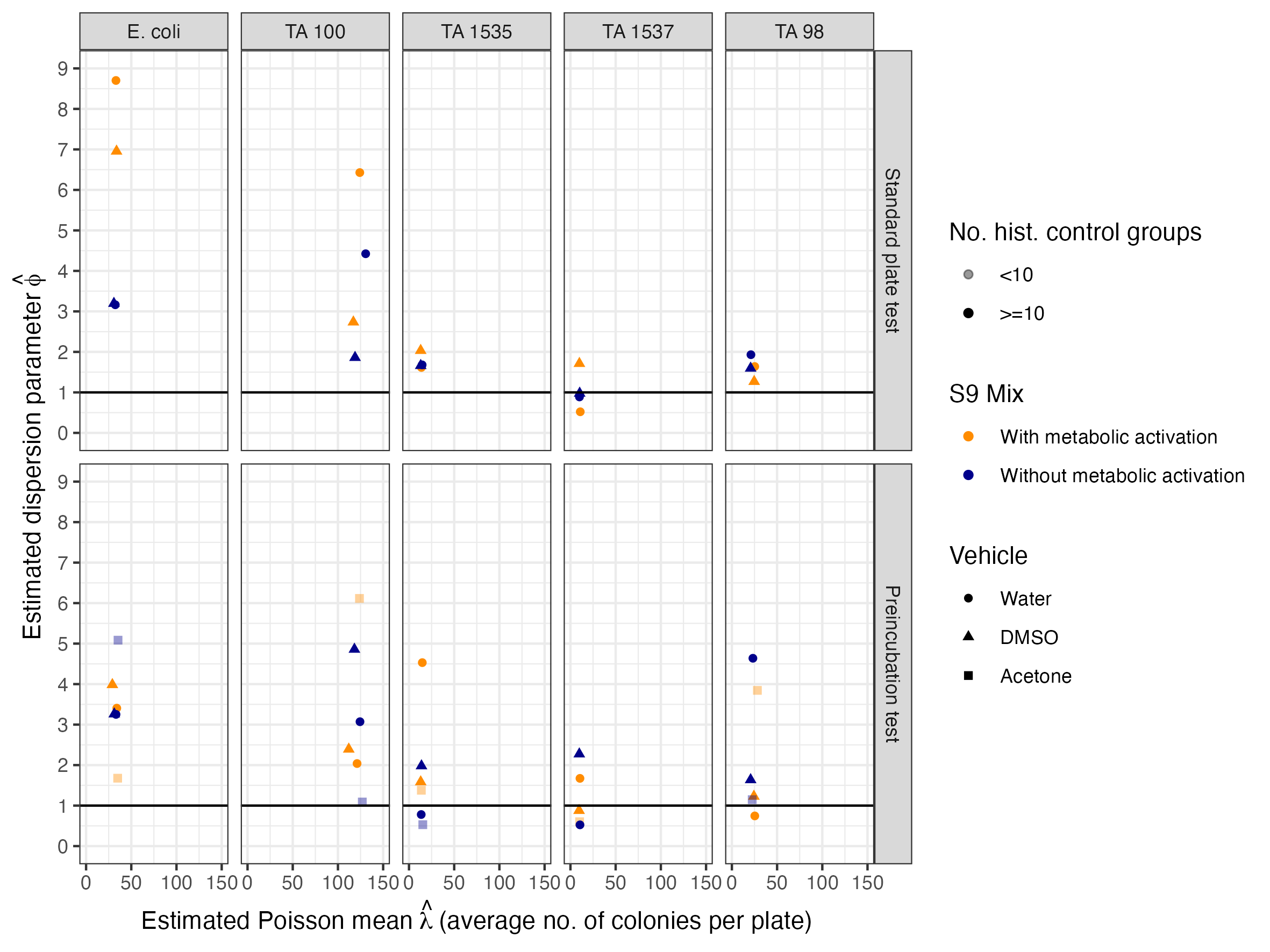}
  \caption{Overview about the estimates for the Poisson mean $\hat{\lambda}$ and the dispersion parameter $\hat{\phi}$ 
  obtained from historical control data with at least five observations per setting.}
  \label{fig:mean_dispersion}
\end{figure}

For a detailed analysis regarding the relationship between the mean and the variance, 
the historical data was split according to the different experimental setups 
(different combinations of bacteria stem, vehicle, S9 mix used or not used and type of assay). 
In order to assess these data sets for possible overdispersion, the model 
described in equations \ref{eq:varQB} and \ref{eq:varBB}, was fitted to each of these 
data sets, if they contained at least five historical control groups (49 out of 90 data sets).
This restriction was used, since the estimate for the dispersion parameter is slightly
negatively biased and can be highly inaccurate, if estimated based on a small number of observations
(McCullagh and Nelder 1989; Menssen and Schaarschidt 2019).
The estimates for the dispersion parameter and the Poisson 
mean for each of these 49 data sets are depicted in figure \ref{fig:mean_dispersion}.

It is apparent that most of the settings can be treated as overdispersed ($\hat{\phi} > 1$).
This means, that in this settings, the observed variability in the data exceeds
the variability that can be modeled by the assumption of a simple Poisson process.\\ 
Especially for the bacteria stems E. coli and TA 100 the data is much more 
variable than under the assumption of simple Poisson distributed data, since the 
estimated dispersion parameter rises up to values above six. The most extreme case of overdispersion
($\hat{\phi} \approx 8.7$) was observed for E. coli in the standard plate test using 
S9 mix and water as the vehicle. Hence, in this data set the observed between
study variability was modeled to be approximately 8.7 times higher, than possible
under the assumption of simple Poisson distributed data.

 
\subsection{Historical data about relapses of multiple sclerosis patients}

\begin{figure}[h!]
  \includegraphics[width=\linewidth]{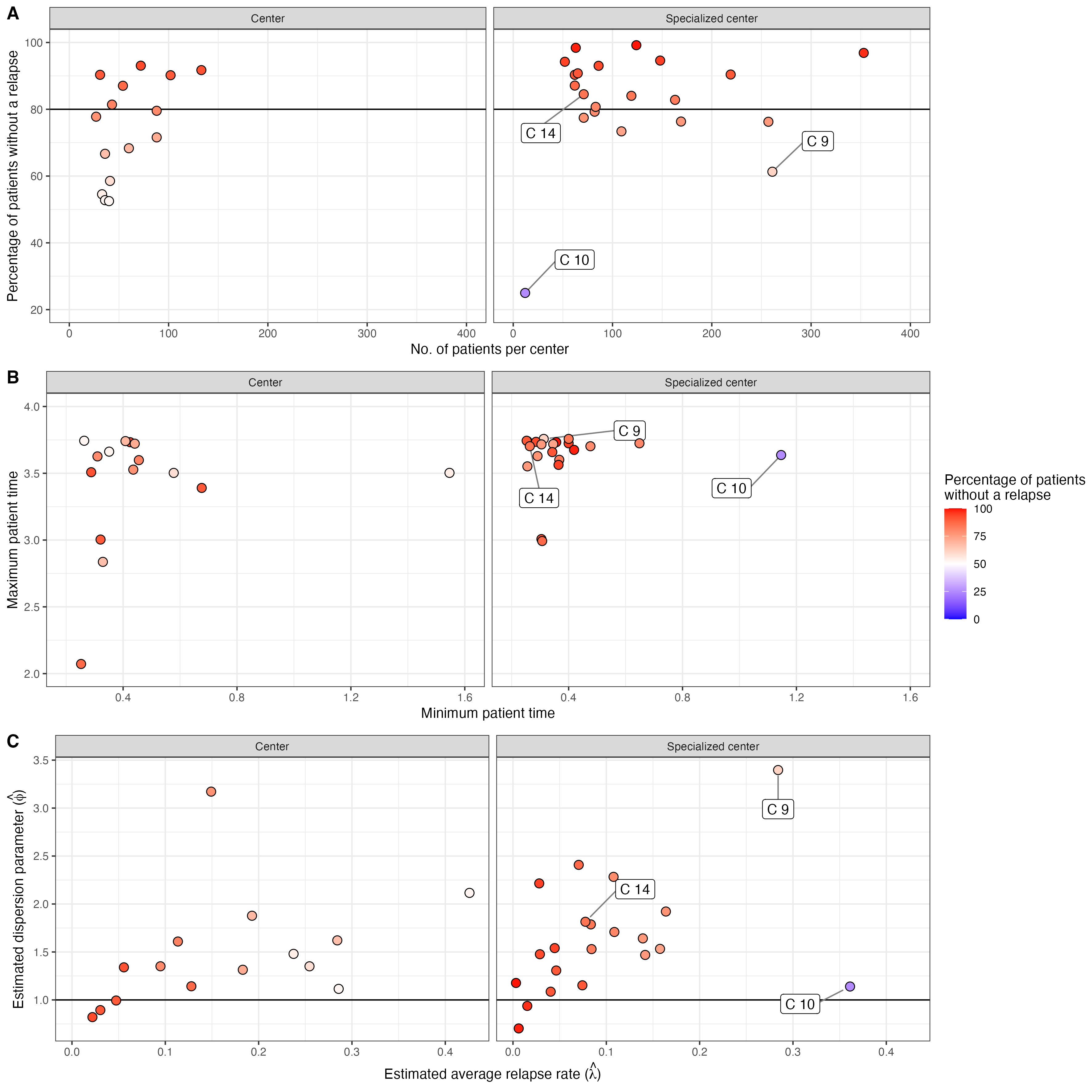}
  \caption{Overview on the data about the number of relapses per multiple sclerosis patient.
  \textbf{A:} Percentages of patients without any relapse per center.
  \textbf{B:} Minimum and maximum time patients are monitored by a certain center.
  \textbf{C:} Estimates for the Poisson mean $\hat{\lambda}$ (the average relapse 
  rate per center and patient year) and the estimated dispersion parameter $\hat{\phi}$ per center.
  Centers 9, 10 and 14 are used in order to demonstrate the application of upper prediction limits (see 
  section \ref{sec::med_control} below).}
  \label{fig:ms_data}
\end{figure}

Longitudinal data on the number of relapses of multiple sclerosis (MS) patients was provided
by the German Multiple Sclerosis Registry (GMSR). This registry accumulates data from 
different types of health care centers (e.g. hospitals or medical practices that work with 
MS patients) across Germany through a certified web-based electronic 
data capture system. A wide range of variables such as demographical
and clinical data are collected. Further details on the data collection of the 
GMSR can be found in Ohle et al. 2021. \\
In this work, the collected data of 36 different centers participating in the 
pharmacovigilance module of the GMSR were assessed in detail, out of which 21 
were specialized centers that predominantly treat MS patients, while the other
15 centres are smaller and less focused on MS in the context of other neurological
diseases. The data set contained observations from 1.1.2020 onwards until the last
visit of each patient. The number of patients per 
center varied between 12 and 353 (fig. 3 A) whereas the observation period of 
those patients varied between  0.25 and 3.76 years (fig. 3 B). The number of 
patients within a center not having a documented relapse during the observation 
period varied between 25\% and 99\%. \\
In order to monitor the average relapse rate $\hat{\lambda}$ per patient per year 
and the between patient overdispersion $\hat{\phi}$ within each center, a GLM that was
based on the quasi-Poisson assumption was fit to each of the 36 data sets (see computational details).
Due to the relatively high number of zeros in some of the data sets, the average relapse rate
per center $\hat{\lambda}$ can become relatively low and varies from 0.0032 to 0.426
(x-axis in figure \ref{fig:ms_data} C). The estimated dispersion parameter per 
center $\hat{\phi}$ varied between 0.70 and 3.39 (y-axis in figure \ref{fig:ms_data} C).
Out of the 36 centers, 31 showed signs of slight to moderate overdispersion 
($\hat{\phi}$ between 1.08 to 2.41), whereas five centers showed signs of underdispersion
($\hat{\phi}$ below 1).


\section{Simulation study} \label{sec:simulation}

The coverage probabilities of the different methods for the calculation of HCL 
reviewed above were assessed by Monte-Carlo simulations with the nominal level 
set to $1-\alpha=0.95$.
The simulation settings were inspired by the real life data shown above and some 
of the parameter combinations used for simulation reflect their properties. Nevertheless,
simulations were run for a broader range of different parameter settings in order
to enable a higher degree of generalization. Simulations were run for both, two 
sided control (or prediction) intervals (reflecting the settings in toxicology) 
as well as for one sided upper prediction bounds (reflecting the data properties 
in medical quality control).\\
For each combination of model parameters, $S=5000$ "historical" data sets were drawn,
on which historical control limits $[l, u]_s$ were calculated. 
Furthermore, $S$ sets of single target values $t^*_s$ (with $t^*=u^*=y^*/n^*$ for 
control limits in u-Charts and $t^*=y^*$ for all other control limits) were sampled and the
coverage probability of the control intervals was computed by 
\begin{gather*}
        \hat{\psi}^{cp}=\frac{\sum_{s=1}^S I_s}{S} \text{ with} \\
        I_s = 1 \text{ if } t^*_s \in [l, u]_s, \\
        I_s = 0 \text{ if } t^*_s \notin [l, u]_s.
\end{gather*}
In order to assess the ability of the different historical control limits to ensure for equal tail 
probabilities (as was required in equation \ref{eq::equal_tails}),
the coverage of the simulated lower borders $l_s$ and upper borders $u_s$ was 
calculated to be
\begin{gather*}
        \hat{\psi}^l=\frac{\sum_{s=1}^S I_s}{S} \text{ with} \\
        I_s = 1 \text{ if } l_s \leq t^*_s, \\
        I_s = 0 \text{ if } l_s > t^*_s.
\end{gather*}
and
\begin{gather}
        \hat{\psi}^u=\frac{\sum_{s=1}^S I_s}{S} \text{ with} \label{eq::psi_u} \\ 
        I_s = 1 \text{ if } t^*_s \leq u_s, \notag \\
        I_s = 0 \text{ if } u_s > t^*_s.\notag
\end{gather}
In the simulations regarding upper prediction limits that served as HCL, their 
coverage probability was calculated according to equation \ref{eq::psi_u}.
Each of the bootstrap-calibrated prediction intervals (or upper limits) that were 
calculated in the simulation was based on $B=10000$ bootstrap-samples. The sampling 
of quasi-Poisson or negative-binomial observations was done following the algorithms
described in section 4 of the supplementary material.


\subsection{Coverage probabilities of two-sided historical control limits}

\begin{figure}[H]
  \includegraphics[width=\linewidth]{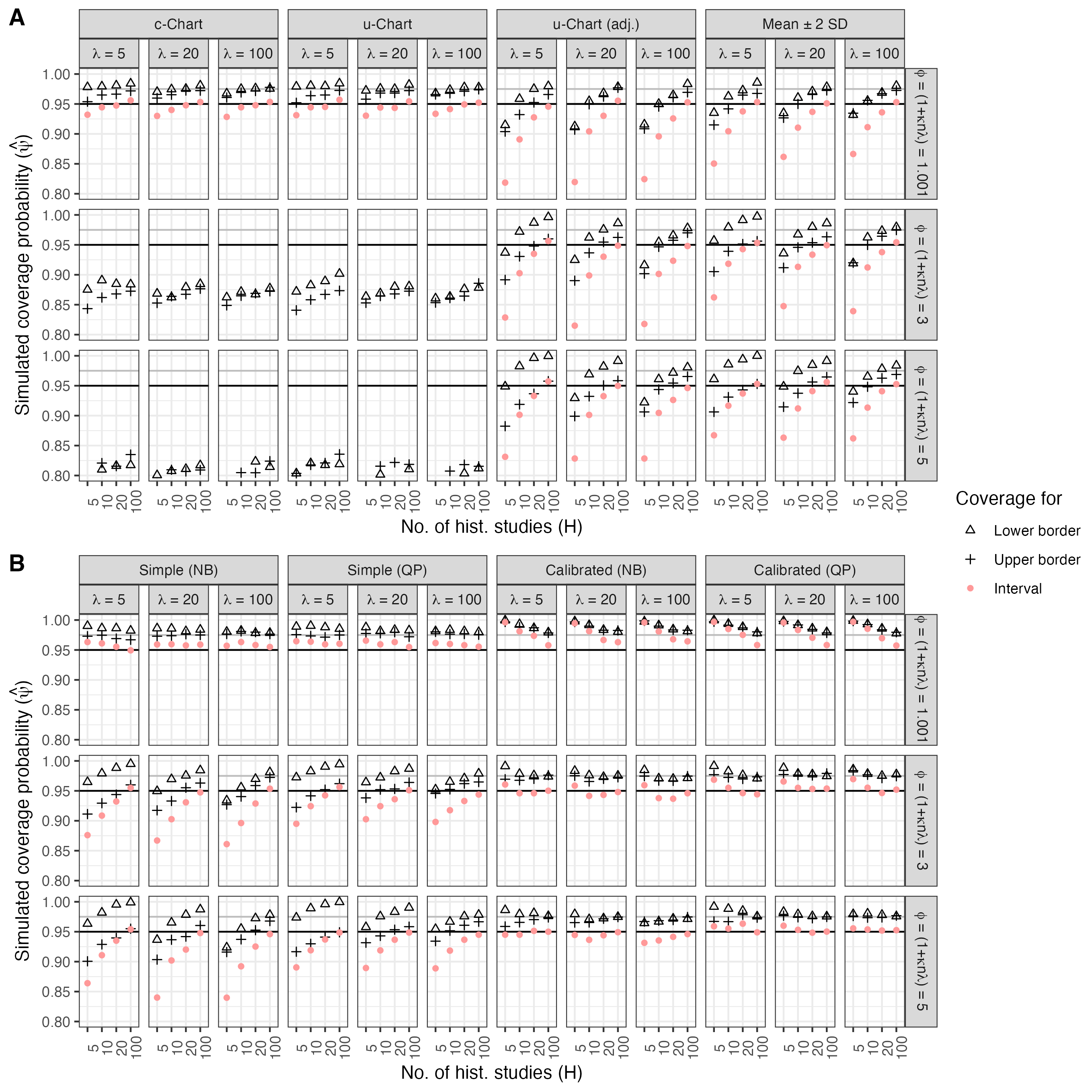}
  \caption{Simulated coverage probabilities based on $n_h=3$ experimental units 
  per control group.
  \textbf{A:} Heuristical control limits.
  \textbf{B:} Prediction intervals.
  $\boldsymbol{\lambda}$: Poisson mean.
  \textbf{Black horizontal lines:} Nominal coverage probability $\psi^{cp}=0.95$. 
  \textbf{Grey horizontal line:} Nominal coverage probability for the lower and 
  the upper limit, if equal tail probabilities are achieved $\psi^{l}=\psi^{u}=0.975$}
  \label{fig:cover_off_3}
\end{figure}

In order to assess the coverage probability $\hat{\psi}^{cp}$ of the two HCL
mentioned above, simulations for all 36 combinations of four different 
numbers of historical clusters $H=\{5, 10, 20, 100\}$ that mimic different numbers
of available historical control groups, three different Poisson means
$\lambda=\{5, 20, 100\}$ and three different dispersion parameters $\phi=\{1.001, 3, 5\}$
were run. Reflecting the experimental design of toxicological studies, these
parameter combinations were used in combination with offsets that remain constant 
between the historical clusters ($n_h=n^*=3$). Hence, in this part of the simulation, 
the mean-variance relationship of the quasi-Poisson assumption is not in contradiction 
with the one of the negative-binomial distribution. \\
The simulated coverage probabilities of the heuristical methods is given in figure 
\ref{fig:cover_off_3} A whereas the coverage probabilities of the prediction intervals
is given in figure \ref{fig:cover_off_3} B. 
The control limits of the c- and u-charts behave similar: Both approach
the nominal coverage probability of 95 \% (red dots) if overdispersion does not
play a role in the data, but show coverage probabilities far below the nominal level 
that decrease down to 58\% in the presence of overdispersion (and hence are not depicted 
in the figure). The adjusted 
u-chart that accounts for overdispersion behaves in a similar way than the control
limits given by the mean $\pm$ 2 standard deviations. 
With a rising number of historical control groups, both methods approach the nominal 
coverage probabilities. But, one has to be aware that both methods do not account for
equal tail probabilities. Hence, their lower limits tend to cover the target value
(one random realization of the data generating process)
almost always, if the Poisson mean is low ($\lambda=5$) and overdispersion plays a role
in the data ($\phi > 1$), because in this case, the underlying distribution is 
heavily right-skewed. \\
The two simple (uncalibrated) prediction intervals (fig. \ref{fig:cover_off_3} B)
tend to behave in a similar 
fashion: They approach the nominal coverage probability if overdispersion plays
only a minor role in the data, but yield coverage below the nominal level, if
overdispersion plays a role and the number of historical control groups is below
20. Since these intervals do not account for equal tail probabilities, one has to
be aware that they practically yield 95\% upper prediction bounds in the case 
of a low Poisson mean and present overdispersion. This is due to the right-skeweness of the
underlying distribution.\\
The calibrated prediction intervals tend to be slightly too conservative, if overdispersion
is absent in the data and the number of historical studies is low. But, if the data is
overdispersed the calibrated prediction intervals approach the nominal coverage 
probabilities, even if computed based on five historical control groups. Since the
calibration algorithm is aimed to adapt the intervals to potential skewness of the underlying 
distribution, the lower and the upper borders of the calibrated prediction intervals
approach the desired 97.5\% coverage probability far closer than any other method.

\subsection{Coverage probabilities of calibrated upper prediction borders}

\begin{figure}[H]
  \includegraphics[width=\linewidth]{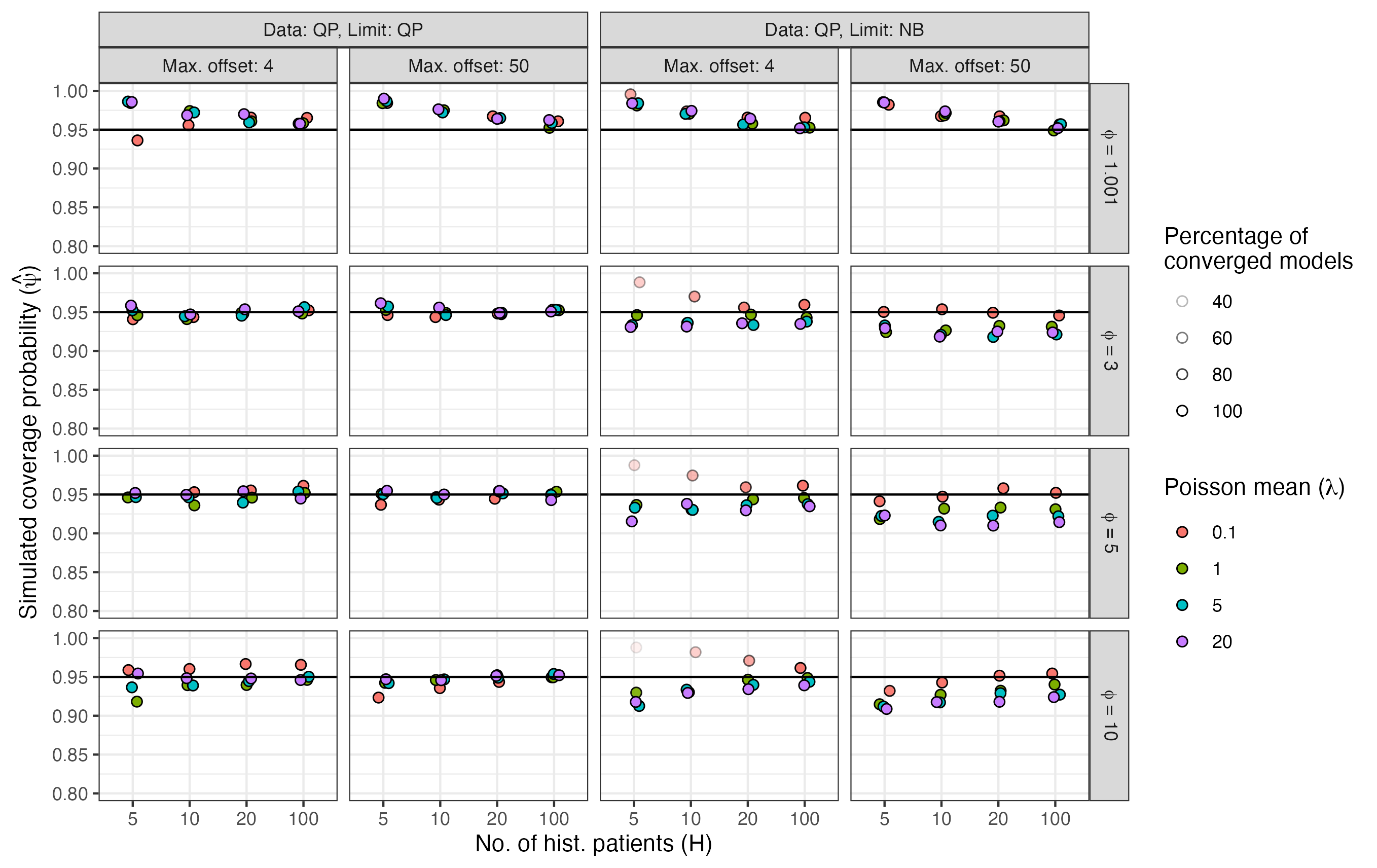}
  \caption{Simulated coverage probabilities of bootstrap-calibrated upper prediction 
  limits based on quasi-Poisson data*.
  $\boldsymbol{\lambda}$: Poisson mean,
  $\boldsymbol{\phi}$: Dispersion parameter,
  \textbf{Max. offset}: Maximum offset used for simulation (the minimum offset was always fixed at 0.5),
  \textbf{Black horizontal lines:} Nominal coverage probability $\psi=0.95$,
  *All coverage probabilities were calculated based on the number of 
  converged model fits.}
  \label{fig:cover_upper_qp}
\end{figure}

\begin{figure}[H]
  \includegraphics[width=\linewidth]{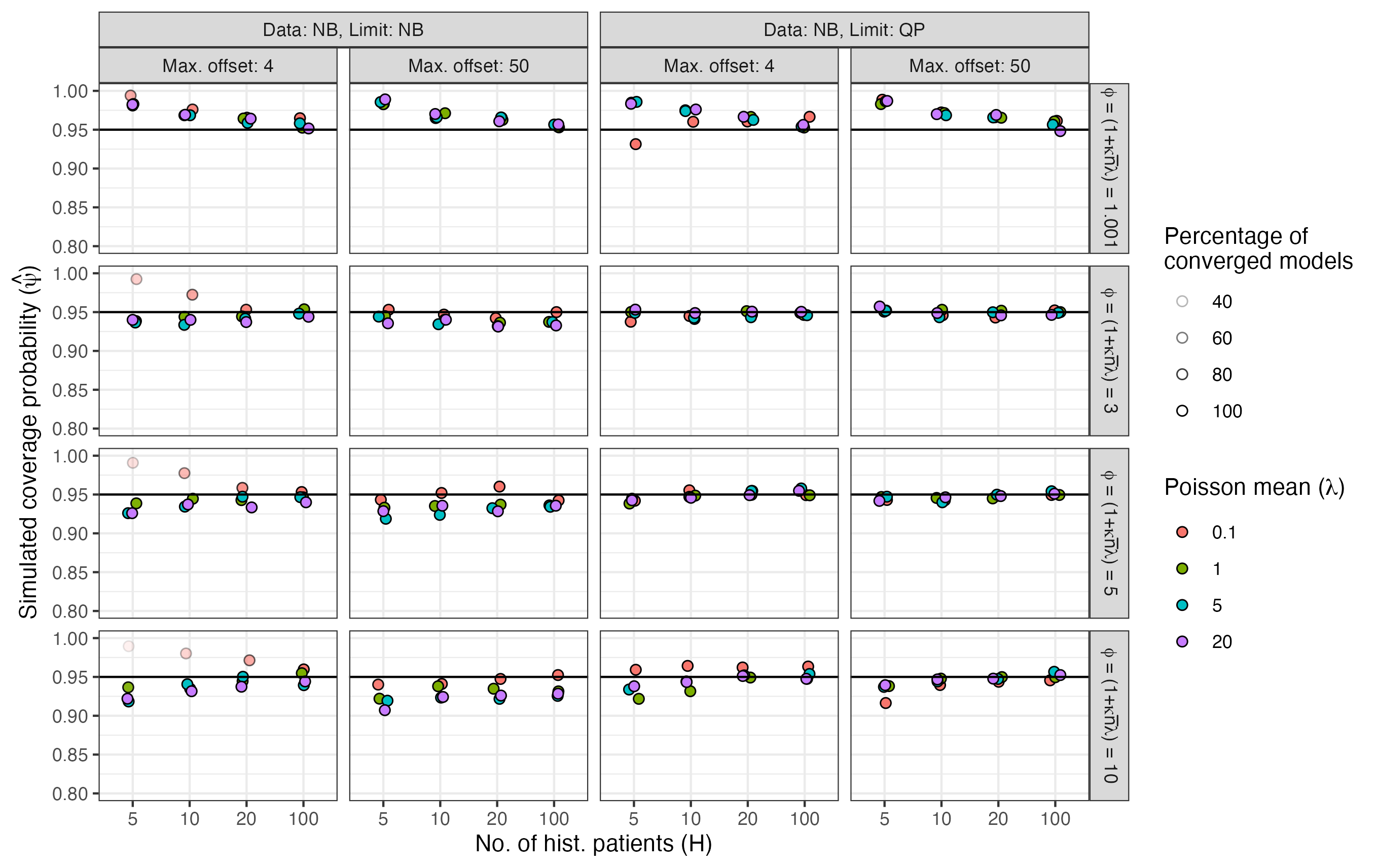}
  \caption{Simulated coverage probabilities of bootstrap-calibrated upper prediction 
  limits based on negative-binomial data*.
  $\boldsymbol{\lambda}$: Poisson mean,
  $\boldsymbol{\phi}$: Dispersion parameter,
  \textbf{Max. offset}: Maximum offset used for simulation (the minimum offset was always fixed at 0.5),
  \textbf{Black horizontal lines:} Nominal coverage probability $\psi=0.95$, 
  *All coverage probabilities were calculated based on the number of 
  converged model fits.}
  \label{fig:cover_upper_nb}
\end{figure}

This part of the simulation was run in order to reflect the application in medical
quality control. Compared to toxicological applications, where the experimental setup
is highly standardized with regard to the experimental design, genetic conditions 
of model organisms etc., medical data is usually derived from less controlled environments.
Patients can differ greatly in age, genetic preposition and other risk factors 
which might lead to high degrees of possible overdispersion. Furthermore, the time 
patients spend under treatment (or in a hospital) can differ greatly between patients,
which has to be taken into account for the application of control limits.
Since medical quality control is usually applied in order to monitor the number of 
adverse events per patient the numbers of counted events is usually relatively low and therefore, 
can contain many zeros. But with an increasing amount of zeros in the data,
the lower limit of control intervals turns out to be less informative
(since it becomes zero itself). Hence, in this scenario
the application of upper control limits has to be favored over the application
of two sided control limits. \\
Since the bootstrap-calibrated prediction intervals outperformed all other 
methodologies in the simulation showed above, only bootstrap-calibrated upper 
prediction bounds are considered in this section.
Monte-Carlo simulations regarding the coverage probability of the calibrated upper 
prediction limits were run based on all combinations of the following parameters
$H=\{5, 10, 20, 100\}$ mimicking the number of patients available for limit calculation,
four different Poisson means $\lambda=\{0.1, 1, 5, 20\}$ and
four different amounts of overdispersion $\phi=\{1.001, 3, 5, 10\}$. \\
In order to reflect the different patient times that were included as an offset 
($n_h$ and $n^*$ in equations \ref{eq::QP_calib} and \ref{eq::NB_calib}), 
offsets were sampled from a uniform distribution. In one step
all 64 parameter combinations were simulated with offsets ranged between 0.5 and 4.
This setting was chosen in order to reflect the patient times in the real life data 
from the GMSR. In order to evaluate the behavior of the prediction limits in the
case were offsets are extremely different, all 64 parameter combinations
were additionally run with offsets drawn from a uniform distribution with minimum
0.5 and maximum 50.\\
Quasi-Poisson data was sampled using the parameter combinations
mentioned above. Based on this type of data, the coverage probabilities of both
calibrated prediction limits (quasi-Poisson and negative-binomial) were assessed.
Vice versa, observations were sampled from the negative-binomial distribution,
to which both methods were applied. This was done in order to assess the 
performance of both methods under model misspecification. \\ 
As noted above, the mean-variance relationship differs between the quasi-Poisson 
assumption and the negative-binomial distribution in the case were offsets are 
different:
In the negative-binomial distribution the Poisson variance is 
inflated by $1 + \kappa n_h \lambda$ (see equation \ref{eq:varBB}). Hence,
contrary to quasi-Poisson data, the magnitude of overdispersion within each 
cluster (eg. patient) depends on the offset and is not a constant anymore.
In order to set the amount of overdispersion in a comparable range as in the 
simulation based on quasi-Poisson data, the parameter $\kappa$ used for data 
sampling was set to $\kappa=\frac{\phi - 1}{\bar{n} \lambda}$ with 
$\bar{n}=\{2.25, 25.25\}$. \\
The simulated coverage probabilities are given in figures \ref{fig:cover_upper_qp}
and \ref{fig:cover_upper_nb}.
Both methods yield coverage probabilities satisfactorily close to the desired 
95\% and seem to be relatively robust against model misspecification.
However, all coverage probabilities are calculated based on the
number simulated data sets to which a fitted model reached convergence.
But, the negative-binomial GLM, fit with \texttt{MASS::glm.nb}, does not 
converge to a given data set in many cases. Especially with a rising 
amount of zeros ($\lambda=0.1$) and declining numbers of available patients 
($H \leq 10$) the number of simulated data sets to which a negative-binomial GLM 
could be fit, became less than 50\% and declined to a minimum of 24.5\%. Contrary to the 
negative-binomial GLM, the GLM based on the quasi-Poisson approach did not show 
any problems with convergence, but yields estimates for the dispersion parameter
$\hat{\phi}$ below one. Due to the necessity that $\hat{\phi}$ was restricted to 
be bigger than 1 in equations \ref{eq::quasi_pois_pi} and \ref{eq::QP_calib}, 
the calibrated quasi-Poisson prediction interval is based on bootstrap data that
was generated based on $\hat{\phi}=1.001$ if the true estimate was below one.

\section{Application of control limits} \label{sec::application}
\subsection{Pre-clinial risk assessment} \label{sec::tarone_application}

In his paper from 1982, Tarone provided the results of a microbial mutagenicity 
assay run to evaluate the mutagenic potential of benz(a)anthracene, alongside with 
HCD that contains 66 control groups from similar mutagenicity assays. Each control
group is comprised of three
petri dishes on which the number of revertant bacteria colonies of the stem TA1537
was counted (tab. 2 and 3 of Tarone 1982). Since the number of petri dishes remains
fixed for all control groups ($n_h=n_{h'}=n^*=3$), one can not distinguish between 
the negative-binomial or quasi-Poisson assumptions. The estimated dispersion 
parameter is $\hat{\phi}=3.18$,
which indicates that the variability of the data generating process from which 
the historical control groups derive is 3.18 times as variable as expected under 
simple Poisson distribution.\\
Different control limits calculated based on the HCD are depicted in table 
\ref{tab::control_limits_tarone}. The control limits from Sheward control charts 
(c- and u-Chart) yield the narrowest control intervals, since they ignore the 
overdispersion present in the HCD. If one multiplies its control limits by $n^*=3$,
the overdispersion adjusted u-chart of Laney (2006) yields control limits that are
comparable to the ones calculated based on the mean $\pm$ 2 SD as well as to the simple 
uncalibrated prediction intervals. This is because all four intervals are symmetrical
and the amount of historical information is relatively high (and hence the uncertainty 
of the estimates used in the heuristical intervals is relatively low). The width 
of the two calibrated prediction intervals is comparable to the width of the uncalibrated 
ones. However, the control limits of the calibrated prediction intervals are 
systematically higher than the limits of the uncalibrated intervals. This is, 
because the calibration accounts for potential skeweness and hence
yields non-symmetrical prediction intervals. 
\begin{table}[h!]
\centering
\caption{Control limits for the number of revertant colonies in three petri dishes ($n^*=3$) 
based on the HCD given by Tarone 1982} 
\begin{tabular}{lrrr}
  \hline
Method & Lower CL & Upper CL & Interval width\\ 
  \hline
  c-Chart & 15.25 (16) & 34.87 (34) & 19.62 \\ 
  u-Chart$^1$ & 5.08 (6) & 11.62 (11) & 6.54 \\ 
  u-Chart (adj.)$^1$ & 2.56 (3) & 14.14 (14) & 11.58 \\ 
  Mean $\pm$ 2 SD & 7.20 (8) & 42.92 (42) & 35.72 \\ 
  Simple (NB)$^2$ & 7.86 (8) & 42.26 (42) & 34.40 \\ 
  Simple (QP)$^3$ & 7.43 (8) & 42.70 (42) & 35.27 \\ 
  Calibrated (NB)$^2$ & 9.90 (10) & 44.67 (44) & 34.76 \\ 
  Calibrated (QP)$^3$ & 9.70 (10) & 45.16 (45) & 35.46 \\ 
   \hline
\end{tabular} \\ [0.5ex]
\raggedright
\scriptsize
\textbf{Numbers in brackets:} Lowest and highest number of counts covered by the interval.\\
\textbf{1:} Control limits of the u-Charts are given for $y^* / n^*$. For control 
limits on the response scale, multiply them by $n^*=3$.\\
\textbf{2:} Estimates $\hat{\lambda}=8.35$ and $\hat{\kappa}=0.082$ were obtained based on \texttt{MASS::glm.nb} (see computational details)\\
\textbf{3:} Estimates $\hat{\lambda}=8.35$ and $\hat{\phi}=3.18$ were obtained based on \texttt{stats::glm} (see computational details)
\label{tab::control_limits_tarone}
\end{table}

\begin{figure}[H]
  \includegraphics[width=\linewidth]{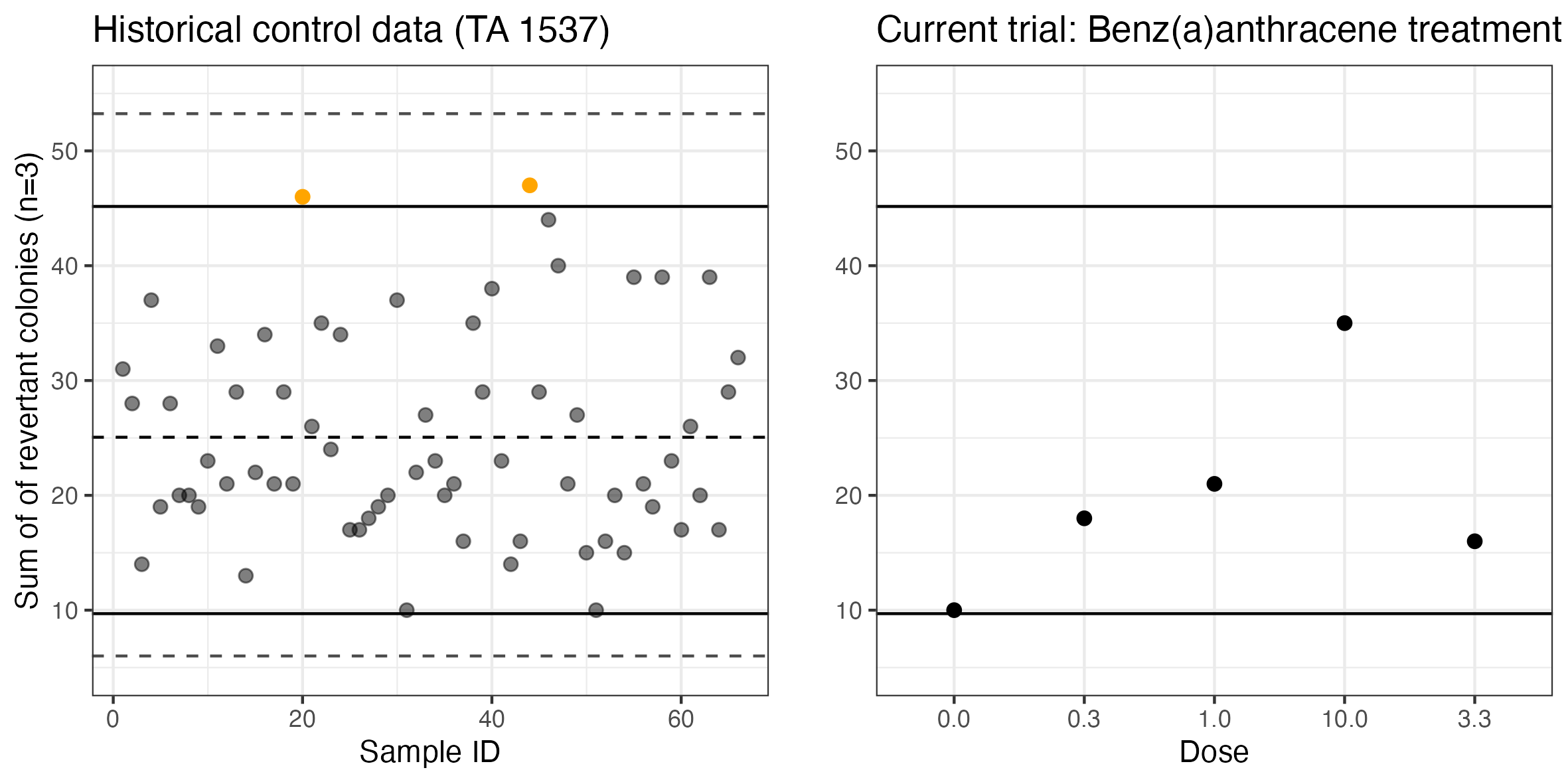}
  \caption{HCD and current trial of Tarone 1982.
  \textbf{Black dashed line:} Expected value for the number of revertant colonies in three petri dishes 
  ($n \hat{\lambda}=25.06$),
  \textbf{Black lines:} Lower and upper border of the 95 \% calibrated quasi-Poisson prediction interval 
  (see tab. \ref{tab::control_limits_tarone}),
  \textbf{Grey dashed lines:} Lower and upper border of a 99 \%  calibrated quasi-Poisson prediction interval,
  \textbf{Orange dots:} Observations that fall outside 95 \% prediction limits.
  Please note that the study ID is artificial.}
  \label{fig:control_chart_tarone}
\end{figure}

Since the calibrated prediction intervals account for the skeweness of the underlying
distribution, they properly approximate the central $100(1-\alpha)$ \% of the underlying
distribution. Hence they can be applied in improved Sheward control charts. The left
panel of figure \ref{fig:control_chart_tarone} shows the HCD (grey dots), the expected number 
of revertant colonies $n\hat{\lambda}=25.06$ (black dashed line), the 95 \% calibrated 
prediction interval (quasi-Poisson) given in table \ref{tab::control_limits_tarone} 
(black lines) and a 99 \% calibrated prediction interval based on the quasi-Poisson
assumption $[l, u]= [6.36, 54.64]$ (grey dashed lines). The right side of figure \ref{fig:control_chart_tarone} 
shows the data of the current trial regarding benz(a)anthracene together with the
95 \% prediction interval. \\
Based on the left panel, one can evaluate the quality of the HCD. Please note, 
that Tarone reported the HCD not as a timeseries, but rather reported the numbers of 
historical control groups in which a certain number of revertant colonies occurred. 
In order to demonstrate how a Sheward control chart would look like if the HCD was reported
as a time series (which is usually the case), each control group was randomly assigned
to an artificial study ID.\\
Based on this artificial study IDs, the process
appears to be constant and most of the historical observations fall into the 95 \% 
prediction interval with two values above the upper limit of the 95 \% interval. 
But, given that there are 66 historical control groups, one can expect $66 * 0.05 = 3.3$
observations outside the central 95 \% (and 0.66 outside the central 99 \%). \\
If one compares the observations from the current trial to the HCD, it is apparent
that the concurrent control group is relatively low, but \textit{all} current 
observations seem to be in line with the historical ones. However, the prediction 
interval applied here has a pointwise interpretation and is explicitly based 
on the assumption, that only the concurrent control group is derived from the
same data generating process as the historical controls. \\
It is important to note, that the evaluation, if the outcome of a complete current
trial (control and treatment groups) is in line with the historical knowledge needs 
further adjustment to account for the multiple testing problem. Furthermore, due
to systematic (but uncontrollable) differences between the different historical 
and current trials, all observations within a trial can be correlated (which causes 
the between study overdispersion). If these correlations play a role, the treatment 
groups can not be treated as independent realizations from the same data generating 
process the control groups are derived from. Hence, control limits that should evaluate 
a whole current trial have to be based on a simultaneous prediction interval, that 
accounts for possible within study correlation. To the authors knowledge, methodology
to compute such a prediction interval is not available so far.\\


\subsection{Medical quality control} \label{sec::med_control}

Contrary to toxicological HCD that usually contains observations which stem from control
groups of controlled experiments run in a consecutive order, data for medical quality control 
is usually gathered on the levels of patients that randomly appear in health care centers. 
Hence, in medical quality control, a given sample of patients which represent the 
observations from a current process is taken as a baseline, to which patients from
another cohort can be compared. \\
Three centers that report their data of MS patients to the GMSR serve as examples:
Center 9, 10 and 14. Center 14 served as the reference based on which estimates for the 
average relapse rate and overdispersion were computed ($\hat{\phi}=1.81$, $\hat{\lambda}=0.077$).
Together with individual patient times, these estimates were used to compute
calibrated 95\% and 99\% upper prediction limits (UPL) for each patient within 
center 14. This was done to identify the patients that showed unusually
high relapse rates and hence, might need clarification, if 
their high number of relapses is explainable by their course of the disease or 
possibly by a reporting error.\\
Compared to center 14, centers 9 and 10 report relatively high relapse rates 
(see fig. \ref{fig:ms_data}) and it is of interest, if both centers report data
that is in line with center 14. Hence, UPL that were based on the estimates for
the average relapse rate $\hat{\lambda}$ and the dispersion parameter $\hat{\phi}$ 
obtained in center 14 and the individual patient times observed in centers 9 and
10 were calculated to which the observed numbers of relapses were compared. \\
Table \ref{tab::ms_UPL} provides an overview about the total number of patients per 
center, the percentage of patients that exceed a certain UPL as well as the corresponding
observed vs. expected number of patients. For
21.45\% of the patients in center 9, the number of relapses exceeds the 95\% UPL
(instead of the expected 5\%). Furthermore, 2.3\% of the patients have higher relapses 
than predicted by the upper 99\% UPL. This clearly indicates, that in center 9 
far more patients exceed the UPL as could be expected, if its underlying data 
generating process was in line with center 14. In other words: The unusual high 
numbers of relapses reported in center 9 can be interpreted as a warning signal, 
that systematic differences between center 14 and center 9 occur that need further
investigation. \\
Similarly as in center 9, also in center 10 more patients than expected exceed the
95\% UPL (3 out of 12), but none of the patients showed numbers of relapses above 
the 99\% UPL. Given the low number of patients in center 10, this can be interpreted
as a weak warning signal that the three patients need further investigation. But,
it remains unclear, if the data generating process in center 10 really differs
from the one in center 14. 
\newpage

\begin{table}[ht]
\centering
\caption{Percentage of patients per center, that exceed pointwise upper prediction limits} 
\begin{tabular}{lrrr}
  \hline
Center & $>$ 95\% UPL & $>$ 99\% UPL & N \\ 
  \hline
C 14 (baseline) & 2.82\%  (2, 3.55) & 1.41\%  (1, 0.71) &  71  \\ 
  C 9 & 21.45\%  (56, 13.05) & 2.30\%  (17, 2.61) &  261 \\ 
  C 10 & 25.00\%  (3,  0.6) & 0.00\%  (0, 0.12) &  12 \\ 
   \hline
\end{tabular}\\ [0.5ex]
\raggedright
\scriptsize
\textbf{Numbers in brackets:} Observed vs. expected numbers of patients with relapses above the control limit.
\textbf{UPL:} Upper prediction limit.
\textbf{N:} Total number of patients per center.
\label{tab::ms_UPL}
\end{table}

\begin{figure}[H]
\includegraphics[width=\linewidth]{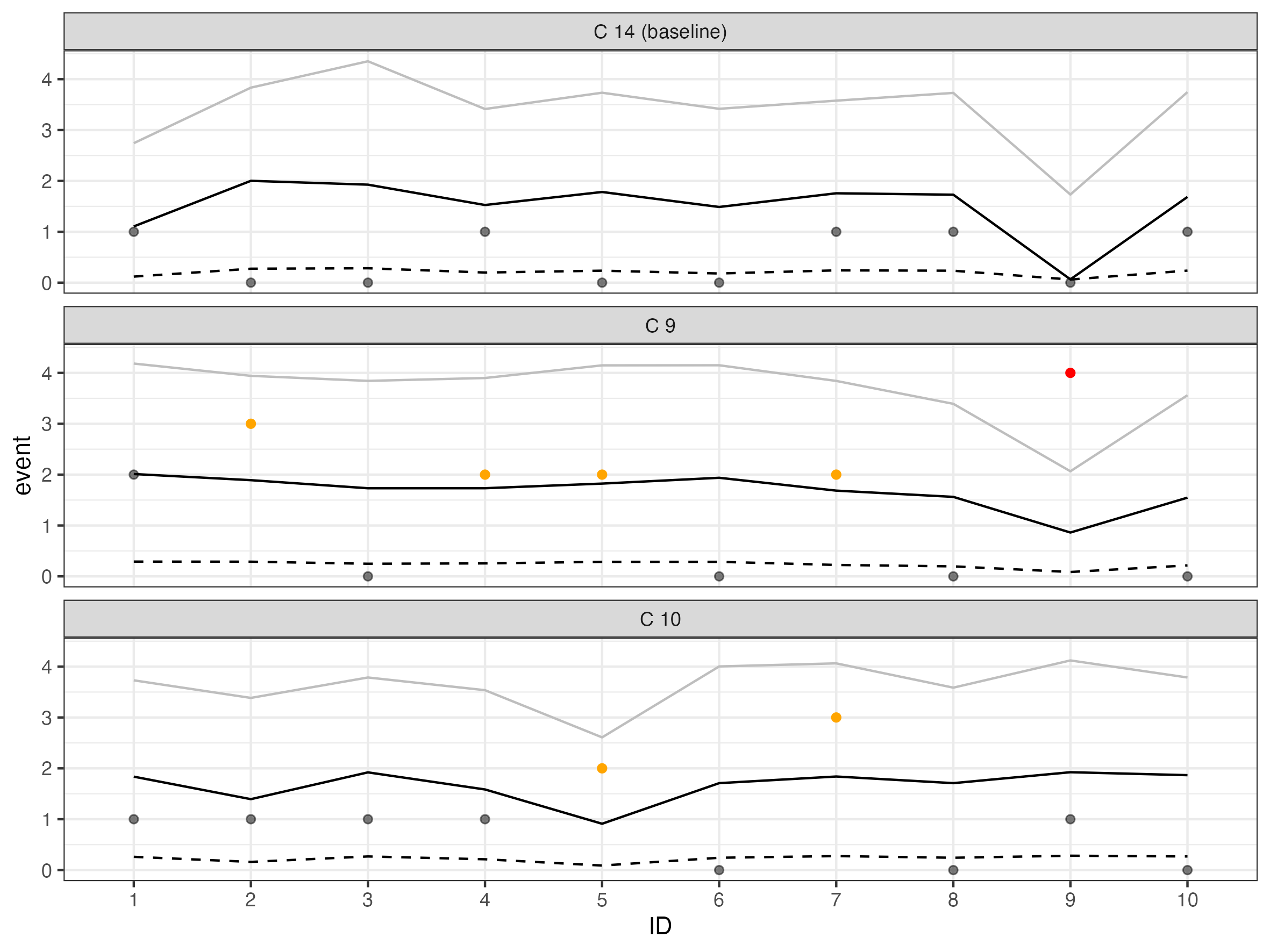}
\caption{Control-chart for the number of relapses for 10 random MS patients per center.
      \textbf{Dashed line}: Expected number of relapses per patient,
      \textbf{Black line:} Calibrated 95 \% upper prediction limits (quasi-Poisson)*,
        \textbf{Grey line:} Calibrated 99 \% upper prediction limits (quasi-Poisson)*,
        \textbf{Grey dots:} Observed number of relapses per patient,
        \textbf{Orange dots:} Observed numbers of relapses per patient exceeds 95 \% upper prediction limit,
        \textbf{Red dots:} Observed numbers of relapses per patient exceeds 95 \% upper prediction limit,
        * All limits are calculated based on estimates ($\hat{\lambda}=0.077$ and 
        $\hat{\phi}=1.81$) obtained from center 14.}
  \label{fig::ms_relapses}
\end{figure}

Sheward type control charts, that are based on the calculated UPL are depicted in 
fig. \ref{fig::ms_relapses}. In order to keep patients unidentifiable, 
10 patients with a maximum of 4 relapses were randomly chosen from each center. \\ 
The expected number of relapses per patient as well as individual 95\% and 99\% UPL are indicated
by the dashed, the black and the grey lines, respectively. The different width 
of the UPL reflects the different times patients have spent under monitoring. Orange 
dots indicate patients above the 95\% UPL whereas red dots indicate patients that
belong to the one percent with the highest relapse rates (given that they would originate 
from the same data generating process as the patients in center 14). As stated above, 
far more patients than expected exceed the UPL in center 9. This can be interpreted 
as a warning signal that the whole data generating process of center 9 differs 
from that in center 14. \\
Beyond that, the Sheward type control chart given in fig. 
\ref{fig::ms_relapses} is a relatively simple tool to detect single patients 
that show an unusual high number of relapses and hence need further investigation:
Since patient 9 of center 9 has spent a relatively short time under monitoring,
also its UPL are relatively low. Consequently, the four relapses of this patient
would belong to the most extreme 1\%, if this patient was prone to the same 
data generating process than the patients in center 14 and hence, needs further 
investigation.

\section{Discussion} \label{sec:discussion}

The analysis of the two real life data bases provides evidence for the presence 
of overdispersion in toxicological and medical count data. The amount of 
overdispersion found in the HCD that descends from the Ames test is in line with 
the findings of others: Obviously the HCD provided by Tarone 1982, that was used 
in section \ref{sec::tarone_application} as an example for the application of the 
proposed methodology, shows clear signs of overdispersion. Furthermore, 
Levy et al. 2019 report HCD from negative control groups about the Ames test that
was provided by more than 20 different laboratories, summarized in 18 different 
data sets. However, they report the HCD in terms of means and standard deviations 
(per data set). If one squares their reported 
standard deviations and compares the resulting variances to the reported means,
the variance exceeds the mean in 12 out of 18 of data sets. With other words, 
also this 12 data sets contain observations of different historical negative 
control groups that show signs of between study overdispersion. \\
The presence of overdispersion in the registry data is in line with medical 
and biopharmaceutical data reported by others: Mohammed et al. 2008 reported a data 
set on the number of falls per patient in a hospital department for which, based 
on the quasi-Poisson assumption, the estimated amount of overdispersion is $\hat{\phi}=1.67$.
Hoffman 2003 reported a data set on bacteria counts in water probes that appeared 
to be heavily overdispersed ($\hat{\phi}=9.38$). \\
This demonstrates the need for methodology that enables the calculation of historical 
control limits for overdispersed count data. Unfortunately the existing methodology
for the calculation of HCL for count data has several drawbacks: All of them are 
based on heuristics that formally lack a clear definition of their statistical
properties (since they lack a formal definition for the desired coverage probability).
Furthermore, they yield HCL symmetrical around the mean, but overdispersed count data
can be heavily right skewed (see fig. 1 in the supplementary material). Consequently all heuristical
methods, reviewed above do not ensure for equal tail probabilities and they 
are not able to approximate the central $100(1-\alpha)$\% of the underlying 
distribution.\\
This gap was closed by the proposed bootstrap calibrated prediction intervals.
Even for a relatively low number of historical clusters (e.g. control groups or 
patients) the bootstrap calibration yields prediction intervals (or limits) with
coverage probabilities close to the nominal level. Since the proposed procedure 
calibrates the lower and the upper limits individually, the resulting prediction 
interval reflects the skeweness of the underlying distribution and hence, ensures 
for equal tail probabilities.\\
Despite the fact that overdispersion is present in toxicological and medical real 
life data and the proposed prediction intervals are able to care about overdispersion,
its presence indicates that some sources of variability are not uniform between 
historical clusters. Especially in toxicological applications, where HCD stems from
controlled experiments, the presence of overdispersion should trigger a closer look
to the historical control data base in order to search for potential systematic 
sources for between study variation that can be controlled. On the other hand 
overdispersion is a common feature of biological count data (McCullagh and Nelder 1989)
because living experimental units can only be standardized up to a certain level 
(e.g. with regard to their genetic condition or age).\\
Therefore the presence of overdispersion might reflect a mixture of controllable and 
uncontrollable sources for between study variability. Due to the possibility, that
between study overdispersion can be caused by a mixture of different sources of
which some might be controllable and others are not, a clear statement on the tolerable 
magnitude of between study overdispersion can not be given here. This must be 
subject to the toxicological research community and needs assay specific 
discussions.\\


\section{Conclusions}

If overdispersion is present in the data (and its magnitude is tolerable):

\begin{itemize}
        \item Common Sheward c- and u-charts do not account for possible right-skeweness
        of the data. Therefore, more observations than desired will fall above
        the upper limit whereas fewer observations than desired will fall below 
        the lower limit.
        \item The bootstrap calibrated prediction intervals yield coverage probabilities 
        close to the nominal level. Furthermore, they account for equal tail probabilities
        and hence, should be favored over all other methods reviewed in this manuscript.
        \item Software for the calculation of bootstrap calibrated prediction intervals is 
        publicly available via the R package predint.
\end{itemize}


\section{References}

Benoit S.W., Goldstein S.L., Dahale D.S., Haslam D.B., Nelson A., Truono K., 
Davies S.M. (2019): Reduction in Nephrotoxic Antimicrobial Exposure Decreases Associated
Acute Kidney Injury in Pediatric Hematopoietic Stem Cell Transplant Patients.
Biology of Blood and Marrow Transplantation 25:1654-1658\\

Chen T-T., Chung K-P., Hu F-C., Fan C-M., Yang M-C. (2010): 
The use of statistical process control (risk-adjusted CUSUM,
risk-adjusted RSPRT and CRAM with prediction limits) for
monitoring the outcomes of out-of-hospital cardiac arrest
patients rescued by the EMS system. Journal of Evaluation in Clinical
 Practice 17:71–77 \\

Coja T., CharistouA., Kyriakopoulou A., Machera K., Mayerhofer U.,
Nikolopoulou D., Spilioti E., Spyropoulou A., Steinwider J., Tripolt T. (2022):
Preparatory work on how to report, use and interpret historical control data in (eco)toxicity
studies. EFSA Supporting Publications 19(9):7558E\\

Demetrio C.G.B., Hinde J. ,Moral R.A. (2014): Models for overdispersed data in entomology.
In: Ferreira C.P., Godoy W.A.C. (eds.) Ecological modelling applied to entomology.
Springer International Publishing 219-259\\

Dertinger S. D., Li D., Beevers C., Douglas G.R., Heflich R.H., 
Lovell D.P., Roberts D.J., Smith R., Uno Y., Williams A., Witt K.L., 
Zeller A., Zhou C. (2023): Assessing the quality and making appropriate use 
of historical negative control data: A report of the International Workshop 
on Genotoxicity Testing (IWGT). Environmental and Molecular Mutagenesis 1-22\\

Deschl U., Kittel B., Rittinghausen S., Morawietz G., Kohler M., Mohr U., Keenan C. (2002):
The Value of Historical Control Data - Scientific Advantages for Pathologists, Industry and Agencies.
Toxicologic Pathology 30(1):80-87\\

EU Comission Regulation 283/2013: Setting out the data requirements for active substances,
in accordance with Regulation (EC) No. 1107/2009 of the European Parliament and of the
 Council concerning the placing of plant protection products on the market.\\

Francq B.G., Lin D., Hoyer W. (2019): Confidence, prediction, and tolerance in 
linear mixed models. Statistics in Medicine 38:5603–5622\\

Greim H., Gelbke H-P., Reuter U., Thielmann H.W., Edler L. (2003): 
Evaluation of historical control data in carcinogenicity studies.
Human and Experimental Toxicology 22:541-549\\

Gsteiger S., Neuenschwander B., Mercier F., Schmidli H. (2013): Using 
historical control information for the design and analysis of clinical trials
with overdispersed count data. Statistics in Medicine 32:3609–3622\\

Gurjanov A., Kreuchwig A., Steiger-Hartmann T., Vaas L.A.I. (2023): 
Hurdles and signposts on the road to virtual control groups—A case
study illustrating the influence of anesthesia protocols on electrolyte 
levels in rats. Fronties in Pharmacology 14:2023\\

Hahn G.J., Meeker W.Q. (1991): Statistical Intervals: A Guide for Practitioners. 
First edition, Wiley NY\\

Hayashi M., Dearfield K., Kasper P., Lovell D., Martus H-J., Thybaud V. (2011):
Compilation and use of genetic toxicity historical control data.
Mutation Research/Genetic Toxicology and Environmental Mutagenesis 723:87–90\\

Hoffman D. (2003): Negative binomial control limits for count data with extra Poisson variation.
Pharmaceutical Statistics 2:127–132\\

Hoffman D., Berger M. (2011): Statistical considerations for calculation of 
immunogenicity screening assay cut points. Journal of Immunological Methods 373:200–208\\

Hothorn L.A. 2015: Statistics in Toxicology Using R. Chapman and Hall\\

Kluxen F.M., Weber K., Strupp C., Jensen S.M., Hothorn L.A., Garcin J-C., Hofmann T. (2021):
Using historical control data in bioassays for regulatory toxicology. 
Regulatory Toxicology and Pharmacology 125:105024\\

Koetsier A., van der Veer S.N., Jager K.J., Peek N., de Keizer N.F. (2012):
Control Charts in Healthcare Quality Improvement. Methods of information on Medicine
51(3):189-198\\

Levy D.D., Zeiger E., Escobar P.A., Hakura A., van der Leede B-J. M., Kato M., 
Moore M.M., Sugiyama K-I. (2019): Recommended criteria for the evaluation of 
bacterial mutagenicity data (Ames test). Mutation Research/Genetic Toxicology 
and Environmental Mutagenesis 848:403074\\

Lyren A., Brilli R.J., Zieker K., Marino M., Muenthing S., Sharek P.J. (2017):
Children’s Hospitals’ Solutions for Patient Safety Collaborative Impact
on Hospital-Acquired Harm. Pediatrics 140(3):e20163494

McCullagh P., Nelder J.A. (1989): Generalized Linear Models. 2nd Edition, 
Chapman and Hall \\

Meeker W.Q., Hahn G.J., Escobar L.A. (2017): Statistical Intervals: A Guide for 
Practitioners and Researchers, 2nd Edition, Wiley\\

Menssen M. 2023: The calculation of historical control limits in toxicology: 
Do's, don'ts and open issues from a statistical perspective. Mutation Research/Genetic
Toxicology and Environmental Mutagenesis 892:503695\\

Menssen M., Schaarschmidt F. (2019): Prediction intervals for overdispersed
 binomial data with application to historical controls. Statistics in Medicine 38(14):2652-2663\\

Menssen M., Schaarschmidt F. (2022): Prediction intervals for all of M future observations 
based on linear random effects models. Statistica Neerlandica 76(3)283-308\\

Mohhamed M.A., Worthington P., Woodall W.H. (2008): Plotting basic control charts: 
tutorial notes for healthcare practitioners. Quality and Safety in Health Care, 17(2):137-4\\

Montgomery D.C. (2020): Introduction to statistical quality control. Wiley, Hoboken NY\\

Nelson W. (1982): Applied Life Data Analysis. Wiley NY\\

NTP (2024): NTP Historical Controls Data Base. https://ntp.niehs.nih.gov/data/controls,
assessed 2.2.2024 \\

OECD (2017): Overview on genetic toxicology TGs, OECD Series on Testing and Assessment, No. 238, OECD Publishing, Paris\\

OECD (471): Test No. 471: Bacterial Reverse Mutation Test. OECD Guidelines for the Testing of
 Chemicals, Section 4, OECD Publishing, Paris\\

OECD (489): Test No. 489: In Vivo Mammalian Alkaline Comet Assay. OECD Guidelines 
for the Testing of Chemicals, Section 4, OECD Publishing, Paris \\

OECD (490): Test No. 490: In Vitro Mammalian Cell Gene Mutation Tests Using the Thymidine Kinase Gene.
OECD Guidelines for the Testing of Chemicals, Section 4, OECD Publishing, Paris\\

Ohle LM., Ellenberger D., Flachenecker P. Friede T., Haas J.,  Hellwig K., 
Parciak T., Warnke C., Paul F., Zettl U.K., Stahlmann A. (2021): Chances and 
challenges of a long-term data repository in multiple sclerosis: 20th birthday 
of the German MS registry. Scientific Reports 11:13340\\

Pognan F., Steger-Hartmann T., Díaz C., Blomberg N., Bringezu F., Briggs K., Callegaro G., Capella-Gutierrez S.,
Centeno E., Corvi J., Drew P., Drewe W.C., Fernández J.M., Furlong L.I., Guney E., Kors J.A., Mayer M.A.,
Pastor M., Piñero J., Ramírez-Anguita J.M., Ronzano F., Rowell P., Saüch-Pitarch J., Valencia A.,
van de Water B., van der Lei J., van Mulligen E., Sanz F. (2021):
The eTRANSAFE Project on Translational Safety Assessment through Integrative Knowledge Management: Achievements and Perspectives. 
Pharmaceuticals (Basel) 14(3):237.\\

Prato E., Biandolino F., Parlapiano I., Grattagliano A., Rotolo F., Buttino i. (2023):
Historical control data of ecotoxicological test with the copepod Tigriopus fulvus.
Chemistry and Ecology 39(8):881-893 \\

Rotolo F., Vitiello V., Pellegrini D., Carotenuto Y., Buttino I. (2021): Historical control
data in ecotoxicology: Eight years of tests with the copepod Acartia tonsa. Environmental Pollution.
284:117468 \\

Sachlas A., Bersimis S., Psarakis S. (2019): Risk-Adjusted Control Charts: Theory, Methods, and
Applications in Health. Statistics in Biosciences 11:630–658\\

Schaarschmidt F., Hofmann M., Jaki T., Gruen B., Hothorn L.A. (2015): Statistical 
approaches for the determination of cut points in anti-drug antibody bioassays.
Journal of Immunological Methodsm 418:84–100\\

Still M.D., Cross L.C., Dunlap M., Rencher R., Larkins E., Carpener D.L., Buchmann T.G., 
Coopersmith C.M. (2013): The Turn Team: A Novel Strategy for Reducing Pressure Ulcers
in the Surgical Intensive Care Unit. Journal of the American College of Surgeons 216(3):373-37 \\

Tejs S. (2008): The Ames test: a methodological short review. Environmental Biotechnology 4(1):7-14\\

Viele, K., Berry, S., Neuenschwander, B., Amzal, B., Chen, F., Enas, N., Hobbs, B., Ibrahim, J.G., 
Kinnersley, N., Lindborg, S., Micallef, S., Roychoudhury, S. and Thompson, L. (2014):
Use of historical control data for assessing treatment effects in clinical trials.
Pharmaceutical Statistics 13:41-54.\\

\end{document}


\begin{titlepage}
\begin{center}

\vspace*{1cm}
\Large{\textbf{}}\\
\vfill
\line(1,0){450}\\
\Large{\textbf{Supplementary materials: Prediction intervals for overdispersed Poisson data and their application in medical and pre-clinical quality control}}\\
\line(1,0){450}\\
\vfill


\end{center}
\end{titlepage}



\setcounter{page}{1}

\section{Modeling with offsets to include baseline quantities} \label{sec::offsets}

A GLM fit to Poisson-type data usually runs on the ln-link (with ln as the natural 
logarithm). Hence it is assumed that the linear predictor is

\begin{equation*}
        ln(\lambda_h) = \eta_h.
\end{equation*}

In such a model, variable baseline quantities $n_h$ can be included as fixed, known 
quantities in the model predictor as an offset (such that no parameter is estimated
for their effect). 
Note that the $n_h$ can contain positive integer or decimal numbers (e.g. different 
numbers of petridishes in $h=1, ... , H$ different control groups or different 
monitoring times of $H$ different patients).
\begin{equation*}
       \eta_h = \beta_0 + ln(n_h) 
\end{equation*}
Due to that step, the model parameter $\beta_0$ is the Poisson mean on the ln-scale
relative to one unit of the offset variable which is shown by the following rearrangement:
\begin{gather*}
	ln(\lambda_h) = \beta_0 + ln(n_h) \\
	ln(\lambda_h) - ln(n_h) = \beta_0 \\
	ln(\frac{\lambda_h}{n_h}) = \beta_0 \\
	exp(\beta_0) = \frac{\lambda_h}{n_h} \\
	exp(\beta_0) n_h =\lambda_h \\
	\lambda n_h = \lambda_h
\end{gather*}
Hence, $\lambda$ is the Poisson mean relative to one unit of the baseline quantitiy $n_h$.



\section{Negative-binomial distribution} \label{sec::neg_binomial_distr}

If the $i= 1, 2, \ldots, n_h$ observations in each of the $h=1, 2, \ldots, H$ historical
clusters are negative-binomial distributed (expressed as a gamma-Poisson mixture)
\begin{equation}
Y_{ih} \sim Pois(\lambda_h)
\end{equation}
with $E(Y_{ih})=\lambda_h$ and 
\begin{equation}
\lambda_h \sim gamma(a, b)
\end{equation}
with $E(\lambda_h) = a/b = \frac{1}{\kappa} / \frac{1}{\kappa \lambda} = \lambda$,
and $var(\lambda_h) = a/b^2 = \frac{1}{\kappa} / \frac{1}{(\kappa \lambda)^2} = \kappa \lambda^2$,
also the sum of the observations per cluster $Y_h=\sum_i^{n_h} Y_{ih}$ is again
\begin{equation}
Y_h \sim Pois(n_h\lambda_h)
\end{equation}
with $E(Y_h)=\sum_i^{n_h} E(\lambda_h) = n_h E(\lambda_h) = n_h \lambda$ and
\begin{equation}
var(Y_h) = n_h \lambda + \kappa (n_h \lambda)^2 = \lambda n_h (1+ \kappa n_h \lambda)
\end{equation}
%


\section{Prediction variances for overdispersed Poisson data} \label{sec:pred_var_deriv}

\subsection{quasi-Poisson prediction variance}

The estimate for the variance of a quasi-Poisson random varianble is
\begin{gather*}
	\widehat{var}(Y_{h})^{QP} =  \hat{\phi} n_h \hat{\lambda}.
\end{gather*}
Hence, the estimate for the variance of the future random variable can be obtained
by replacing $n_h$ by $n^*$
\begin{gather*}
	\widehat{var}(Y^*)^{QP} =  \hat{\phi} n^* \hat{\lambda}.
\end{gather*}
If $\hat{\Lambda}=\bar{n} \hat{\lambda}$ with $\bar{n}=\frac{\sum_h^H n_h}{H}$,
then
\begin{gather*}
	\widehat{var}(\hat{\Lambda})^{QP} = \widehat{var}(\bar{n} \hat{\lambda})^{QP} =
	\frac{\hat{\phi} \hat{\Lambda}}{H} = \frac{\hat{\phi} \bar{n} \hat{\lambda}}{H}
\end{gather*}
and 
\begin{gather*}
	\widehat{var}(\hat{\lambda})^{QP} = \widehat{var}(\hat{\Lambda} / \bar{n})^{QP} =
	\frac{1}{\bar{n}^2} \widehat{var}(\hat{\Lambda})^{QP} =
	\frac{\hat{\phi} \hat{\lambda}}{\bar{n} H}.
\end{gather*}
Hence,
\begin{gather*}
	\widehat{var}(n^* \hat{\lambda})^{QP}= n^{*2} \frac{\hat{\phi} \hat{\lambda}}{\bar{n} H}
\end{gather*}
and 
\begin{gather*}
	\widehat{var}(n^* \hat{\lambda} - Y^* )^{QP}=
	\widehat{var}(n^* \hat{\lambda})^{QP} + \widehat{var}(Y^*)^{QP} =
	n^{*2} \frac{\hat{\phi} \hat{\lambda}}{\bar{n} H} +
	\hat{\phi} n^* \hat{\lambda} 
\end{gather*}
%


\subsection{Negative-binomial prediction variance}

The estimate for the variance of a negative-binomial random varianble is
\begin{gather*}
	\widehat{var}(Y_h)^{NB}=n_h \hat{\lambda} + \hat{\kappa} n_h^2 \hat{\lambda}^2 = 
	n_h \hat{\lambda} (1+ \hat{\kappa} n_h \hat{\lambda}). 
\end{gather*}
Hence, the estimate for the variance of the future random variable can be obtained
by replacing $n_h$ by $n^*$
\begin{gather*}
	\widehat{var}(Y^*)^{NB}=n^*  \hat{\lambda} + \hat{\kappa} n^{*2}  \hat{\lambda}^2 = 
	n^*  \hat{\lambda} (1+ \hat{\kappa} n^*  \hat{\lambda}). 
\end{gather*}
If $\hat{\Lambda}=\bar{n} \hat{\lambda}$ with $\bar{n}=\frac{\sum_h^H n_h}{H}$,
then
\begin{gather*}
	\widehat{var}(\hat{\Lambda})^{NB} = \widehat{var}(\bar{n}\hat{\lambda})^{NB} =
	\frac{\hat{\Lambda} + \hat{\kappa} \hat{\Lambda}^2}{H} =
	\frac{\bar{n} \hat{\lambda} + \hat{\kappa} (\bar{n}\hat{\lambda})^2}{H}
\end{gather*}
and
\begin{gather*}
	\widehat{var}(\hat{\lambda})^{NB} = 
	\widehat{var}(\hat{\Lambda} / \bar{n})^{NB} =
	\frac{1}{\bar{n}^2} \widehat{var}(\hat{\lambda})^{NB} =
	\frac{\hat{\lambda} + \hat{\kappa} \bar{n} \hat{\lambda}^2}{\bar{n}H}.
\end{gather*}
Hence,
\begin{gather*}
	\widehat{var}(n^* \hat{\lambda})^{NB}= n^{*2} \frac{\hat{\lambda} + \hat{\kappa} \bar{n} \hat{\lambda}^2}{\bar{n}H}
\end{gather*}
and 
\begin{gather*}
	\widehat{var}(n^* \hat{\lambda} - Y^* )^{NB}=
	\widehat{var}(n^* \hat{\lambda})^{NB} + \widehat{var}(Y^*)^{NB} =
	n^{*2} \frac{\hat{\lambda} + \hat{\kappa} \bar{n} \hat{\lambda}^2}{\bar{n}H} +
	n^* \lambda (1+ \kappa n^* \lambda).
\end{gather*}
%


\section{Sampling of overdispersed Poisson type data} \label{sec:sampling}

\subsection{Sampling of quasi-Poisson data}
Observations with variance
\begin{equation*}
        var(Y_i)=n_i \lambda (1+ \kappa_i n_i \lambda ) = \phi n_i \lambda
\end{equation*}
that descent from $i=1, 2, \ldots, I$ clusters can be sampled from the 
negative-binomial distribution based on predefined values for
the dispersion parameter $\phi$, the number of experimental units per cluster
$n_i$ and the overall Poisson mean $\lambda$ using the following algorithm:\\
For each cluster define $\kappa_i$ as
\begin{equation*}
        \kappa_i = \frac{\phi - 1}{n_i \lambda}.
\end{equation*}
with $\phi > 1$.
Then calculate the parameters of the corresponding gamma distributions as
$a_i=\frac{1}{\kappa_i}$ and $b_i = \frac{1}{\kappa_i n_i \lambda}$ and sample
the Poisson means $\lambda_i$ for each cluster, such that
\begin{equation*}
        \lambda_i \sim Gamma(a_i, b_i).
\end{equation*}
Finally, the observations $y_i$ are sampled from the Poisson distribution
\begin{equation*}
        y_i \sim Pois(\lambda_i)
\end{equation*}
This sampling algorithm is implemented in the function \texttt{rqpois()} of the
R package predint.

\subsection{Sampling of negative-binomial data}

Observations that descent from the negative-binomial distribution have variance
\begin{equation*}
        var(Y_i) = n_i \lambda (1+ \kappa n_i \lambda).
\end{equation*}
Negative-biomial observations can be sampled based on predefined values of $\kappa$, 
$\lambda$ and $n_i$: \\
Define the parameters of the gamma distribution as $a=\frac{1}{\kappa}$ and
$b_i=\frac{1}{\kappa n_i \lambda}$. Then, sample the Poisson means for each cluster
\begin{equation*}
        \lambda_i \sim Gamma(a, b_i).
\end{equation*}
Finally, the observations $y_i$ are sampled from the Poisson distribution
\begin{equation*}
        y_i \sim Pois(\lambda_i)
\end{equation*}
This sampling algorithm is implemented in the function \texttt{rnbinom()} of the
R package predint.


\section{Visual overview about overdispersed Poisson data} \label{sec:histograms}

\begin{figure}[h!]
  \includegraphics[width=\linewidth]{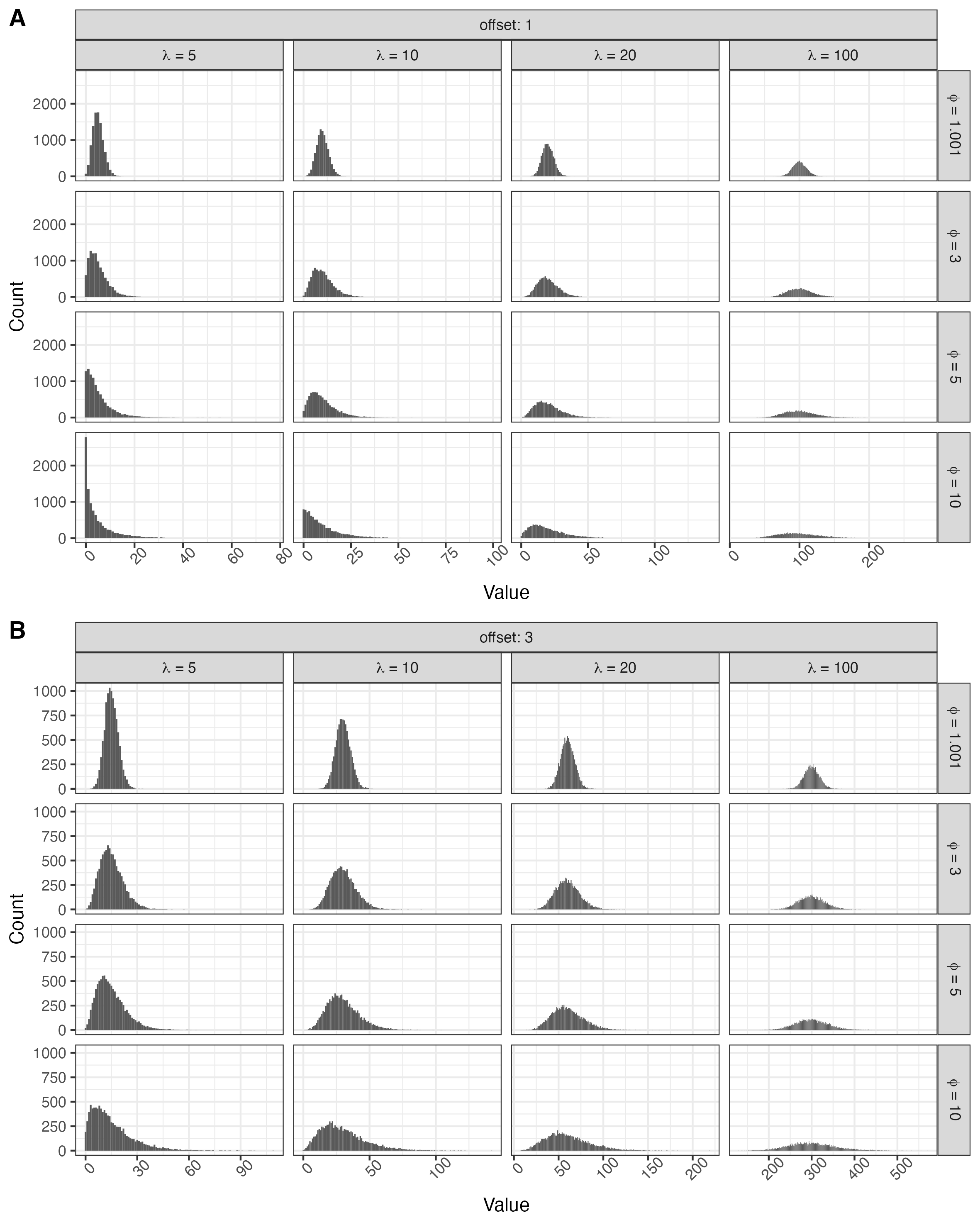}
  \caption{Histograms of 10000 observations, each sampled based on the parameter
  settings of the simulation study.
  \textbf{A:} Offset $n_h=1$, \textbf{B:} Offset $n_h=3$}
  \label{fig:histograms}
\end{figure}

The properties of the data, the simulation study depends on (see section 5 of the manuscript),
are shownin figure \ref{fig:histograms}. For this purpose, 10000 observations were sampled 
for each of the combinations of the mean $\lambda$ and the dispersion parameter $\phi$
for both offset variables ($n_h=1$ and $n_h=3$). It is important to note, that the 
right-skewness of the underlying distribution is rising, the closer the mean goes 
to zero and / or the higher the dispersion parameter is. 

